\documentclass[lettersize,journal]{IEEEtran}

\def\BibTeX{{\rm B\kern-.05em{\sc i\kern-.025em b}\kern-.08em
    T\kern-.1667em\lower.7ex\hbox{E}\kern-.125emX}}

\usepackage{silence}
\usepackage[hyphens]{url}
\usepackage[breaklinks,colorlinks]{hyperref}
\usepackage[usenames,dvipsnames]{xcolor}
\hypersetup{citecolor=blue,linkcolor=blue}
\usepackage[backend=biber, style=ieee, doi=false,isbn=false,giveninits=true,maxbibnames=99,sorting=none]{biblatex}
\usepackage{fancyhdr,lastpage}
\usepackage{epsfig}
\usepackage{balance} 
\usepackage{color}
\usepackage{comment}
\usepackage{epic}
\usepackage{epsf}
\usepackage{listings}
\usepackage{makecell}
\usepackage{multirow}
\usepackage{textcomp}
\usepackage{xspace}    
\usepackage{latexsym}
\usepackage[labelfont=bf,font=small,skip=5pt]{caption}
\usepackage{afterpage}
\usepackage{amstext}   
\usepackage{amsmath,amsfonts}
\usepackage{graphicx}
\usepackage{tabularx}
\usepackage[caption=false,font=normalsize,labelfont=sf,textfont=sf]{subfig}
\usepackage{longtable}
\usepackage{rotating}
\usepackage[algoruled, linesnumbered,noend]{algorithm2e}
\usepackage{fp}
\usepackage{siunitx}
\usepackage{xstring}
\usepackage{multirow}
\usepackage{tabularx}
\usepackage{ragged2e}
\usepackage{amssymb}
\usepackage{booktabs}

\usepackage{calculator}
\usepackage{calculus}
\usepackage{siunitx} 
\usepackage{numprint}

\usepackage{enumitem}

\usepackage{threeparttable}
\usepackage{pifont}

\usepackage{xpatch}
\xpatchbibdriver{online}
  {\printtext[parens]{\usebibmacro{date}}}
  {\iffieldundef{year}{}{\printtext[parens]{\usebibmacro{date}}}}
  {}{}  

\definecolor{bblue}{HTML}{4F81BD}
\definecolor{rred}{HTML}{C0504D}
\definecolor{ggreen}{HTML}{9BBB59}
\definecolor{ppurple}{HTML}{9F4C7C}
\definecolor{darkgray}{rgb}{0.66, 0.66, 0.66}
\definecolor{gray}{RGB}{136,136,136}
\definecolor{dkgreen}{rgb}{0,0.6,0}
\definecolor{gray}{rgb}{0.5,0.5,0.5}
\definecolor{mauve}{rgb}{0.58,0,0.82}
\definecolor{comment-red}{rgb}{0.8,0,0}

\newcommand{\Norothead}[2][0]{\makebox[9mm][c]{\rotatebox{#1}{\makecell[c]{#2}}}}
\newlength\tbspace
\newcolumntype{C}{c<{\hspace{\tbspace}}}

\SetCommentSty{mycommfont}

\makeatletter
\newcommand\footnoteref[1]{\protected@xdef\@thefnmark{\ref{#1}}\@footnotemark}
\makeatother
\newcommand{\quotes}[1]{``#1''}
\newcommand{\DefMacro}[2]{\expandafter\newcommand\csname rmk-#1\endcsname{#2}}
\newcommand{\UseMacro}[1]{\csname rmk-#1\endcsname}

\newcommand{\Space}[1]{}

\makeatletter
\newcommand{\labitem}[2]{%
\def\@itemlabel{\textbf{#1}}
\item
\def\@currentlabel{#1}\label{#2}}

\SetKwProg{Fn}{Function}{}{}

\newcommand{\PP}[1]{
\vspace{2px}
\noindent{\bf \IfEndWith{#1}{.}{#1}{#1.}}
}
\renewcommand{\paragraph}[1]{\smallskip\noindent\emph{#1}\quad}

\newcommand{\V}{\pmb{\checkmark}}
\newcommand{\Vc}{\pmb{\checkmark\kern-1.1ex\raisebox{.7ex}{\rotatebox[origin=c]{125}{--}}}}
\newcommand{\xmark}{\ding{55}}%

\newcommand{\Sys}{CPD\xspace} 
\newcommand{\Edge}{pseudo-edge\xspace}
\newcommand{\Edges}{pseudo-edges\xspace}
\newcommand{\EDges}{Pseudo-edges\xspace}

\newcommand{\bigEI}[1]{$\mathcal{E}_{#1}$\xspace}
\newcommand{\bigPI}[1]{$\mathcal{P}_{#1}$\xspace}
\newcommand{\bigLI}[1]{$\mathcal{L}_{#1}$\xspace}
\newcommand{\bigTI}[1]{$\mathcal{T}_{#1}$\xspace}
\newcommand{\bigNI}[1]{$\mathcal{N}_{#1}$\xspace}

\SetCommentSty{algocommfont}

\usepackage{tikz}

\newcommand\submittedtext{%
  \footnotesize This work has been submitted to the IEEE for possible publication. Copyright may be transferred without notice, after which this version may no longer be accessible.}

\newcommand\submittednotice{%
\begin{tikzpicture}[remember picture,overlay]
\node[anchor=south,yshift=10pt] at (current page.south) {\fbox{\parbox{\dimexpr0.65\textwidth-\fboxsep-\fboxrule\relax}{\submittedtext}}};
\end{tikzpicture}%
}

\DefMacro{ATLASv2_COMMANDER_Stage1_FP}{1602}
\DefMacro{ATLASv2_COMMANDER_Stage1_FN}{0} 
\DefMacro{ATLASv2_COMMANDER_Stage2_FP}{11}
\DefMacro{ATLASv2_COMMANDER_Stage2_FN}{0} 
\DefMacro{ATLASv2_COMMANDER_Stage3_FN}{0} 
\DefMacro{ATLASv2_CBC_FP}{56} 
\DefMacro{ATLASv2_CBC_FN}{0} 
 
\DefMacro{DARPA_E5_COMMANDER_Stage3_FP}{7}
\DefMacro{DARPA_E5_COMMANDER_Stage3_FN}{0}

\DefMacro{DARPA_E5_FD1_COMMANDER_Stage1_FP}{3361}
\DefMacro{DARPA_E5_FD1_COMMANDER_Stage1_FN}{0}
\DefMacro{DARPA_E5_FD1_COMMANDER_Stage2_FP}{105}
\DefMacro{DARPA_E5_FD1_COMMANDER_Stage2_FN}{0}
\DefMacro{DARPA_E5_FD1_Elastic_FP}{1595} 
\DefMacro{DARPA_E5_FD1_Elastic_FN}{0} 
\DefMacro{DARPA_E5_FD1_Sigma_FP}{13317}
\DefMacro{DARPA_E5_FD1_Sigma_FN}{0}

\DefMacro{DARPA_E5_FD2_COMMANDER_Stage1_FP}{7771}
\DefMacro{DARPA_E5_FD2_COMMANDER_Stage1_FN}{0}
\DefMacro{DARPA_E5_FD2_COMMANDER_Stage2_FP}{12}
\DefMacro{DARPA_E5_FD2_COMMANDER_Stage2_FN}{0}
\DefMacro{DARPA_E5_FD2_Elastic_FP}{2301} 
\DefMacro{DARPA_E5_FD2_Elastic_FN}{1} 
\DefMacro{DARPA_E5_FD2_Sigma_FP}{27551}
\DefMacro{DARPA_E5_FD2_Sigma_FN}{0}

\DefMacro{DARPA_E5_FD3_COMMANDER_Stage1_FP}{10779}
\DefMacro{DARPA_E5_FD3_COMMANDER_Stage1_FN}{0}
\DefMacro{DARPA_E5_FD3_COMMANDER_Stage2_FP}{0}
\DefMacro{DARPA_E5_FD3_COMMANDER_Stage2_FN}{0}
\DefMacro{DARPA_E5_FD3_Elastic_FP}{1475} 
\DefMacro{DARPA_E5_FD3_Elastic_FN}{0} 
\DefMacro{DARPA_E5_FD3_Sigma_FP}{17001}
\DefMacro{DARPA_E5_FD3_Sigma_FN}{0}

\DefMacro{DARPA_OpTC_COMMANDER_Stage1_FP}{63460}
\DefMacro{DARPA_OpTC_COMMANDER_Stage1_FN}{0}
\DefMacro{DARPA_OpTC_COMMANDER_Stage2_FP}{35}
\DefMacro{DARPA_OpTC_COMMANDER_Stage2_FN}{0}
\DefMacro{DARPA_OpTC_COMMANDER_Stage3_FP}{4}
\DefMacro{DARPA_OpTC_COMMANDER_Stage3_FN}{0}
\DefMacro{DARPA_OpTC_Elastic_FP}{15111}
\DefMacro{DARPA_OpTC_Elastic_FN}{1} 
\DefMacro{DARPA_OpTC_Chronicle_FP}{47584}
\DefMacro{DARPA_OpTC_Chronicle_FN}{0}
\DefMacro{DARPA_OpTC_Sigma_FP}{11567}
\DefMacro{DARPA_OpTC_Sigma_FN}{0}

\DefMacro{APT29_Scenario1_COMMANDER_Stage1_FP}{1640}
\DefMacro{APT29_Scenario1_COMMANDER_Stage2_FP}{18} 
\DefMacro{APT29_Scenario1_COMMANDER_Stage3_FP}{23}
\DefMacro{APT29_Scenario1_COMMANDER_Stage3_FN}{0}

\DefMacro{APT29_Scenario1_g3mef76_malicious_COMMANDER_Stage1_FP}{1595}
\DefMacro{APT29_Scenario1_g3mef76_malicious_COMMANDER_Stage1_FN}{0}
\DefMacro{APT29_Scenario1_g3mef76_malicious_COMMANDER_Stage2_FP}{23}
\DefMacro{APT29_Scenario1_g3mef76_malicious_COMMANDER_Stage2_FN}{0}
\DefMacro{APT29_Scenario1_g3mef76_malicious_Elastic_FP}{54}
\DefMacro{APT29_Scenario1_g3mef76_malicious_Elastic_FN}{2}
\DefMacro{APT29_Scenario1_g3mef76_malicious_Chronicle_FP}{560}
\DefMacro{APT29_Scenario1_g3mef76_malicious_Chronicle_FN}{1}
\DefMacro{APT29_Scenario1_g3mef76_malicious_Sigma_FP}{53}
\DefMacro{APT29_Scenario1_g3mef76_malicious_Sigma_FN}{0}
\DefMacro{APT29_Scenario1_g3mef76_benign_COMMANDER_Stage1_FP}{1788}
\DefMacro{APT29_Scenario1_g3mef76_benign_COMMANDER_Stage1_FN}{0}
\DefMacro{APT29_Scenario1_g3mef76_benign_COMMANDER_Stage2_FP}{23}
\DefMacro{APT29_Scenario1_g3mef76_benign_COMMANDER_Stage2_FN}{0}
\DefMacro{APT29_Scenario1_g3mef76_benign_Elastic_FP}{71}
\DefMacro{APT29_Scenario1_g3mef76_benign_Elastic_FN}{0}
\DefMacro{APT29_Scenario1_g3mef76_benign_Chronicle_FP}{488}
\DefMacro{APT29_Scenario1_g3mef76_benign_Chronicle_FN}{0}
\DefMacro{APT29_Scenario1_g3mef76_benign_Sigma_FP}{77}
\DefMacro{APT29_Scenario1_g3mef76_benign_Sigma_FN}{0}

\DefMacro{APT29_Scenario1_g3mef77_malicious_COMMANDER_Stage1_FP}{36}
\DefMacro{APT29_Scenario1_g3mef77_malicious_COMMANDER_Stage1_FN}{0}
\DefMacro{APT29_Scenario1_g3mef77_malicious_COMMANDER_Stage2_FP}{1}
\DefMacro{APT29_Scenario1_g3mef77_malicious_COMMANDER_Stage2_FN}{0}
\DefMacro{APT29_Scenario1_g3mef77_malicious_Elastic_FP}{5}
\DefMacro{APT29_Scenario1_g3mef77_malicious_Elastic_FN}{0}
\DefMacro{APT29_Scenario1_g3mef77_malicious_Chronicle_FP}{48}
\DefMacro{APT29_Scenario1_g3mef77_malicious_Chronicle_FN}{0}
\DefMacro{APT29_Scenario1_g3mef77_malicious_Sigma_FP}{2}
\DefMacro{APT29_Scenario1_g3mef77_malicious_Sigma_FN}{0}
\DefMacro{APT29_Scenario1_g3mef77_benign_COMMANDER_Stage1_FP}{1488}
\DefMacro{APT29_Scenario1_g3mef77_benign_COMMANDER_Stage1_FN}{0}
\DefMacro{APT29_Scenario1_g3mef77_benign_COMMANDER_Stage2_FP}{16}
\DefMacro{APT29_Scenario1_g3mef77_benign_COMMANDER_Stage2_FN}{0}
\DefMacro{APT29_Scenario1_g3mef77_benign_Elastic_FP}{53}
\DefMacro{APT29_Scenario1_g3mef77_benign_Elastic_FN}{0}
\DefMacro{APT29_Scenario1_g3mef77_benign_Chronicle_FP}{352}
\DefMacro{APT29_Scenario1_g3mef77_benign_Chronicle_FN}{0}
\DefMacro{APT29_Scenario1_g3mef77_benign_Sigma_FP}{77}
\DefMacro{APT29_Scenario1_g3mef77_benign_Sigma_FN}{0}

\DefMacro{APT29_Scenario1_rc1r6g07ik7_malicious_COMMANDER_Stage1_FP}{97}
\DefMacro{APT29_Scenario1_rc1r6g07ik7_malicious_COMMANDER_Stage1_FN}{0}
\DefMacro{APT29_Scenario1_rc1r6g07ik7_malicious_COMMANDER_Stage2_FP}{0}
\DefMacro{APT29_Scenario1_rc1r6g07ik7_malicious_COMMANDER_Stage2_FN}{0}
\DefMacro{APT29_Scenario1_rc1r6g07ik7_malicious_Elastic_FP}{29}
\DefMacro{APT29_Scenario1_rc1r6g07ik7_malicious_Elastic_FN}{0}
\DefMacro{APT29_Scenario1_rc1r6g07ik7_malicious_Chronicle_FP}{8}
\DefMacro{APT29_Scenario1_rc1r6g07ik7_malicious_Chronicle_FN}{0}
\DefMacro{APT29_Scenario1_rc1r6g07ik7_malicious_Sigma_FP}{2724}
\DefMacro{APT29_Scenario1_rc1r6g07ik7_malicious_Sigma_FN}{0} 
\DefMacro{APT29_Scenario1_rc1r6g07ik7_benign_COMMANDER_Stage1_FP}{1009}
\DefMacro{APT29_Scenario1_rc1r6g07ik7_benign_COMMANDER_Stage1_FN}{0}
\DefMacro{APT29_Scenario1_rc1r6g07ik7_benign_COMMANDER_Stage2_FP}{19}
\DefMacro{APT29_Scenario1_rc1r6g07ik7_benign_COMMANDER_Stage2_FN}{0}
\DefMacro{APT29_Scenario1_rc1r6g07ik7_benign_Elastic_FP}{48}
\DefMacro{APT29_Scenario1_rc1r6g07ik7_benign_Elastic_FN}{0}
\DefMacro{APT29_Scenario1_rc1r6g07ik7_benign_Chronicle_FP}{224}
\DefMacro{APT29_Scenario1_rc1r6g07ik7_benign_Chronicle_FN}{0}
\DefMacro{APT29_Scenario1_rc1r6g07ik7_benign_Sigma_FP}{12225}
\DefMacro{APT29_Scenario1_rc1r6g07ik7_benign_Sigma_FN}{0}

\DefMacro{APT29_Scenario2_COMMANDER_Stage3_FP}{14} 
\DefMacro{APT29_Scenario2_COMMANDER_Stage3_FN}{0} 

\DefMacro{APT29_Scenario2_g3mef76_malicious_COMMANDER_Stage1_FP}{594}
\DefMacro{APT29_Scenario2_g3mef76_malicious_COMMANDER_Stage1_FN}{0}
\DefMacro{APT29_Scenario2_g3mef76_malicious_COMMANDER_Stage2_FP}{9}
\DefMacro{APT29_Scenario2_g3mef76_malicious_COMMANDER_Stage2_FN}{0}
\DefMacro{APT29_Scenario2_g3mef76_malicious_Elastic_FP}{15}
\DefMacro{APT29_Scenario2_g3mef76_malicious_Elastic_FN}{1}
\DefMacro{APT29_Scenario2_g3mef76_malicious_Chronicle_FP}{212}
\DefMacro{APT29_Scenario2_g3mef76_malicious_Chronicle_FN}{1}
\DefMacro{APT29_Scenario2_g3mef76_malicious_Sigma_FP}{3}
\DefMacro{APT29_Scenario2_g3mef76_malicious_Sigma_FN}{0}
\DefMacro{APT29_Scenario2_g3mef76_benign_COMMANDER_Stage1_FP}{1543}
\DefMacro{APT29_Scenario2_g3mef76_benign_COMMANDER_Stage1_FN}{0}
\DefMacro{APT29_Scenario2_g3mef76_benign_COMMANDER_Stage2_FP}{12}
\DefMacro{APT29_Scenario2_g3mef76_benign_COMMANDER_Stage2_FN}{0}
\DefMacro{APT29_Scenario2_g3mef76_benign_Elastic_FP}{88}
\DefMacro{APT29_Scenario2_g3mef76_benign_Elastic_FN}{0}
\DefMacro{APT29_Scenario2_g3mef76_benign_Chronicle_FP}{816}
\DefMacro{APT29_Scenario2_g3mef76_benign_Chronicle_FN}{0}
\DefMacro{APT29_Scenario2_g3mef76_benign_Sigma_FP}{34}
\DefMacro{APT29_Scenario2_g3mef76_benign_Sigma_FN}{0}

\DefMacro{APT29_Scenario2_g3mef77_malicious_COMMANDER_Stage1_FP}{121}
\DefMacro{APT29_Scenario2_g3mef77_malicious_COMMANDER_Stage1_FN}{0}
\DefMacro{APT29_Scenario2_g3mef77_malicious_COMMANDER_Stage2_FP}{0}
\DefMacro{APT29_Scenario2_g3mef77_malicious_COMMANDER_Stage2_FN}{0}
\DefMacro{APT29_Scenario2_g3mef77_malicious_Elastic_FP}{5}
\DefMacro{APT29_Scenario2_g3mef77_malicious_Elastic_FN}{1}
\DefMacro{APT29_Scenario2_g3mef77_malicious_Chronicle_FP}{52}
\DefMacro{APT29_Scenario2_g3mef77_malicious_Chronicle_FN}{0}
\DefMacro{APT29_Scenario2_g3mef77_malicious_Sigma_FP}{2}
\DefMacro{APT29_Scenario2_g3mef77_malicious_Sigma_FN}{0}
\DefMacro{APT29_Scenario2_g3mef77_benign_COMMANDER_Stage1_FP}{942}
\DefMacro{APT29_Scenario2_g3mef77_benign_COMMANDER_Stage1_FN}{0}
\DefMacro{APT29_Scenario2_g3mef77_benign_COMMANDER_Stage2_FP}{8}
\DefMacro{APT29_Scenario2_g3mef77_benign_COMMANDER_Stage2_FN}{0}
\DefMacro{APT29_Scenario2_g3mef77_benign_Elastic_FP}{66}
\DefMacro{APT29_Scenario2_g3mef77_benign_Elastic_FN}{0}
\DefMacro{APT29_Scenario2_g3mef77_benign_Chronicle_FP}{444}
\DefMacro{APT29_Scenario2_g3mef77_benign_Chronicle_FN}{0}
\DefMacro{APT29_Scenario2_g3mef77_benign_Sigma_FP}{35}
\DefMacro{APT29_Scenario2_g3mef77_benign_Sigma_FN}{0}

\DefMacro{APT29_Scenario2_rc1r6g07ik7_malicious_COMMANDER_Stage1_FP}{1}
\DefMacro{APT29_Scenario2_rc1r6g07ik7_malicious_COMMANDER_Stage1_FN}{0}
\DefMacro{APT29_Scenario2_rc1r6g07ik7_malicious_COMMANDER_Stage2_FP}{0}
\DefMacro{APT29_Scenario2_rc1r6g07ik7_malicious_COMMANDER_Stage2_FN}{0}
\DefMacro{APT29_Scenario2_rc1r6g07ik7_malicious_Elastic_FP}{2}
\DefMacro{APT29_Scenario2_rc1r6g07ik7_malicious_Elastic_FN}{0}
\DefMacro{APT29_Scenario2_rc1r6g07ik7_malicious_Chronicle_FP}{0}
\DefMacro{APT29_Scenario2_rc1r6g07ik7_malicious_Chronicle_FN}{0}
\DefMacro{APT29_Scenario2_rc1r6g07ik7_malicious_Sigma_FP}{2428}
\DefMacro{APT29_Scenario2_rc1r6g07ik7_malicious_Sigma_FN}{0}
\DefMacro{APT29_Scenario2_rc1r6g07ik7_benign_COMMANDER_Stage1_FP}{1118}
\DefMacro{APT29_Scenario2_rc1r6g07ik7_benign_COMMANDER_Stage1_FN}{0}
\DefMacro{APT29_Scenario2_rc1r6g07ik7_benign_COMMANDER_Stage2_FP}{33}
\DefMacro{APT29_Scenario2_rc1r6g07ik7_benign_COMMANDER_Stage2_FN}{0}
\DefMacro{APT29_Scenario2_rc1r6g07ik7_benign_Elastic_FP}{175}
\DefMacro{APT29_Scenario2_rc1r6g07ik7_benign_Elastic_FN}{0}
\DefMacro{APT29_Scenario2_rc1r6g07ik7_benign_Chronicle_FP}{200}
\DefMacro{APT29_Scenario2_rc1r6g07ik7_benign_Chronicle_FN}{0}
\DefMacro{APT29_Scenario2_rc1r6g07ik7_benign_Sigma_FP}{10803}
\DefMacro{APT29_Scenario2_rc1r6g07ik7_benign_Sigma_FN}{0}

\DefMacro{Sandworm_Scenario1_g3mef76_malicious_COMMANDER_Stage1_FP}{865}
\DefMacro{Sandworm_Scenario1_g3mef76_malicious_COMMANDER_Stage1_FN}{0}
\DefMacro{Sandworm_Scenario1_g3mef76_malicious_COMMANDER_Stage2_FP}{3}
\DefMacro{Sandworm_Scenario1_g3mef76_malicious_COMMANDER_Stage2_FN}{0}
\DefMacro{Sandworm_Scenario1_g3mef76_malicious_Elastic_FP}{10}
\DefMacro{Sandworm_Scenario1_g3mef76_malicious_Elastic_FN}{1}
\DefMacro{Sandworm_Scenario1_g3mef76_malicious_Chronicle_FP}{225}
\DefMacro{Sandworm_Scenario1_g3mef76_malicious_Chronicle_FN}{0}
\DefMacro{Sandworm_Scenario1_g3mef76_malicious_Sigma_FP}{2}
\DefMacro{Sandworm_Scenario1_g3mef76_malicious_Sigma_FN}{0}
\DefMacro{Sandworm_Scenario1_g3mef76_benign_COMMANDER_Stage1_FP}{1062}
\DefMacro{Sandworm_Scenario1_g3mef76_benign_COMMANDER_Stage1_FN}{0}
\DefMacro{Sandworm_Scenario1_g3mef76_benign_COMMANDER_Stage2_FP}{1}
\DefMacro{Sandworm_Scenario1_g3mef76_benign_COMMANDER_Stage2_FN}{0}
\DefMacro{Sandworm_Scenario1_g3mef76_benign_Elastic_FP}{186}
\DefMacro{Sandworm_Scenario1_g3mef76_benign_Elastic_FN}{0}
\DefMacro{Sandworm_Scenario1_g3mef76_benign_Chronicle_FP}{472}
\DefMacro{Sandworm_Scenario1_g3mef76_benign_Chronicle_FN}{0}
\DefMacro{Sandworm_Scenario1_g3mef76_benign_Sigma_FP}{0}
\DefMacro{Sandworm_Scenario1_g3mef76_benign_Sigma_FN}{0}

\DefMacro{Sandworm_Scenario1_g3mef77_malicious_COMMANDER_Stage1_FP}{74}
\DefMacro{Sandworm_Scenario1_g3mef77_malicious_COMMANDER_Stage1_FN}{0}
\DefMacro{Sandworm_Scenario1_g3mef77_malicious_COMMANDER_Stage2_FP}{0}
\DefMacro{Sandworm_Scenario1_g3mef77_malicious_COMMANDER_Stage2_FN}{0}
\DefMacro{Sandworm_Scenario1_g3mef77_malicious_Elastic_FP}{4}
\DefMacro{Sandworm_Scenario1_g3mef77_malicious_Elastic_FN}{0}
\DefMacro{Sandworm_Scenario1_g3mef77_malicious_Chronicle_FP}{48}
\DefMacro{Sandworm_Scenario1_g3mef77_malicious_Chronicle_FN}{0}
\DefMacro{Sandworm_Scenario1_g3mef77_malicious_Sigma_FP}{0}
\DefMacro{Sandworm_Scenario1_g3mef77_malicious_Sigma_FN}{0}
\DefMacro{Sandworm_Scenario1_g3mef77_benign_COMMANDER_Stage1_FP}{840}
\DefMacro{Sandworm_Scenario1_g3mef77_benign_COMMANDER_Stage1_FN}{0}
\DefMacro{Sandworm_Scenario1_g3mef77_benign_COMMANDER_Stage2_FP}{5}
\DefMacro{Sandworm_Scenario1_g3mef77_benign_COMMANDER_Stage2_FN}{0}
\DefMacro{Sandworm_Scenario1_g3mef77_benign_Elastic_FP}{227}
\DefMacro{Sandworm_Scenario1_g3mef77_benign_Elastic_FN}{0}
\DefMacro{Sandworm_Scenario1_g3mef77_benign_Chronicle_FP}{376}
\DefMacro{Sandworm_Scenario1_g3mef77_benign_Chronicle_FN}{0}
\DefMacro{Sandworm_Scenario1_g3mef77_benign_Sigma_FP}{0}
\DefMacro{Sandworm_Scenario1_g3mef77_benign_Sigma_FN}{0}

\DefMacro{Sandworm_Scenario1_rc1r6g07ik7_malicious_COMMANDER_Stage1_FP}{45}
\DefMacro{Sandworm_Scenario1_rc1r6g07ik7_malicious_COMMANDER_Stage1_FN}{0}
\DefMacro{Sandworm_Scenario1_rc1r6g07ik7_malicious_COMMANDER_Stage2_FP}{0}
\DefMacro{Sandworm_Scenario1_rc1r6g07ik7_malicious_COMMANDER_Stage2_FN}{0}
\DefMacro{Sandworm_Scenario1_rc1r6g07ik7_malicious_Elastic_FP}{14}
\DefMacro{Sandworm_Scenario1_rc1r6g07ik7_malicious_Elastic_FN}{0}
\DefMacro{Sandworm_Scenario1_rc1r6g07ik7_malicious_Chronicle_FP}{0}
\DefMacro{Sandworm_Scenario1_rc1r6g07ik7_malicious_Chronicle_FN}{0}
\DefMacro{Sandworm_Scenario1_rc1r6g07ik7_malicious_Sigma_FP}{1542}
\DefMacro{Sandworm_Scenario1_rc1r6g07ik7_malicious_Sigma_FN}{0}
\DefMacro{Sandworm_Scenario1_rc1r6g07ik7_benign_COMMANDER_Stage1_FP}{336}
\DefMacro{Sandworm_Scenario1_rc1r6g07ik7_benign_COMMANDER_Stage1_FN}{0}
\DefMacro{Sandworm_Scenario1_rc1r6g07ik7_benign_COMMANDER_Stage2_FP}{12}
\DefMacro{Sandworm_Scenario1_rc1r6g07ik7_benign_COMMANDER_Stage2_FN}{0}
\DefMacro{Sandworm_Scenario1_rc1r6g07ik7_benign_Elastic_FP}{86}
\DefMacro{Sandworm_Scenario1_rc1r6g07ik7_benign_Elastic_FN}{0}
\DefMacro{Sandworm_Scenario1_rc1r6g07ik7_benign_Chronicle_FP}{88}
\DefMacro{Sandworm_Scenario1_rc1r6g07ik7_benign_Chronicle_FN}{0}
\DefMacro{Sandworm_Scenario1_rc1r6g07ik7_benign_Sigma_FP}{62216}
\DefMacro{Sandworm_Scenario1_rc1r6g07ik7_benign_Sigma_FN}{0}

\DefMacro{Sandworm_Scenario1_Win_COMMANDER_Stage3_FP}{3} 
\DefMacro{Sandworm_Scenario1_Win_COMMANDER_Stage3_FN}{0} 

\DefMacro{Sandworm_Scenario1_malicious_Linux_COMMANDER_Stage1_FP}{439}
\DefMacro{Sandworm_Scenario1_malicious_Linux_COMMANDER_Stage1_FN}{0}
\DefMacro{Sandworm_Scenario1_malicious_Linux_COMMANDER_Stage2_FP}{13}
\DefMacro{Sandworm_Scenario1_malicious_Linux_COMMANDER_Stage2_FN}{0}
\DefMacro{Sandworm_Scenario1_malicious_Linux_COMMANDER_Stage3_FP}{3}
\DefMacro{Sandworm_Scenario1_malicious_Linux_COMMANDER_Stage3_FN}{3}

\DefMacro{Sandworm_Scenario1_benign_Linux_COMMANDER_Stage1_FP}{1134}
\DefMacro{Sandworm_Scenario1_benign_Linux_COMMANDER_Stage1_FN}{0}
\DefMacro{Sandworm_Scenario1_benign_Linux_COMMANDER_Stage2_FP}{14}
\DefMacro{Sandworm_Scenario1_benign_Linux_COMMANDER_Stage2_FN}{0}
\DefMacro{Sandworm_Scenario1_benign_Linux_COMMANDER_Stage3_FP}{3}
\DefMacro{Sandworm_Scenario1_benign_Linux_COMMANDER_Stage3_FN}{0}

\DefMacro{Sandworm_Scenario1_Linux_COMMANDER_Stage3_FP}{5}
\DefMacro{Sandworm_Scenario1_Linux_COMMANDER_Stage3_FN}{0}

\DefMacro{ATLASv2_Memory_Usage_max}{579}
\DefMacro{DARPA_E5_Memory_Usage_max}{18464}
\DefMacro{DARPA_OpTC_Memory_Usage_max}{10830}
\DefMacro{APT29_Scenario1_Memory_Usage_max}{3625}
\DefMacro{APT29_Scenario2_Memory_Usage_max}{3513}
\DefMacro{Sandworm_Scenario1_Memory_Usage_max}{3484}

\DefMacro{ATLASv2_Memory_Usage_mean}{247}
\DefMacro{DARPA_E5_Memory_Usage_mean}{11720}
\DefMacro{DARPA_OpTC_Memory_Usage_mean}{4484}
\DefMacro{APT29_Scenario1_Memory_Usage_mean}{892}
\DefMacro{APT29_Scenario2_Memory_Usage_mean}{895}
\DefMacro{Sandworm_Scenario1_Memory_Usage_mean}{573}

\DefMacro{fpr_avg_reduction}{93}
\DefMacro{fpr_avg_2to3_reduction}{83}
\DefMacro{fp_reduction_CBC}{80}

\begin{document}

\title{Accurate and Scalable Detection and Investigation of Cyber Persistence Threats}

\author{Qi Liu, Muhammad Shoaib, Mati Ur Rehman, Kaibin Bao, Veit Hagenmeyer, Wajih Ul Hassan
\thanks{This work was supported by funding of the Helmholtz Association (HGF) through the Energy System Design (ESD) program. We also acknowledge support by the Karlsruhe House of Young Scientists (KHYS) for the research stay of Qi Liu at University of Virginia.

Qi Liu, Kaibin Bao, and Veit Hagenmeyer are with Institute for Automation and Applied Informatics, Karlsruhe Institute of Technology (KIT), Eggenstein-Leopoldshafen 76344, Germany (e-mail: qi.liu@kit.edu; kaibin.bao@kit.edu; veit.hagenmeyer@kit.edu).

Muhammad Shoaib, Mati Ur Rehman, and Wajih Ul Hassan are with the School of Engineering \& Applied Science, University of Virginia, Charlottesville, VA 22904-4740, USA (e-mail: mshoaib@virginia.edu; wkw9be@virginia.edu; hassan@virginia.edu).
}}

\maketitle

\submittednotice 

\begin{abstract}

    In Advanced Persistent Threat (APT) attacks, achieving stealthy persistence within target systems is often crucial for an attacker's success. This persistence allows adversaries to maintain prolonged access, often evading detection mechanisms. Recognizing its pivotal role in the APT lifecycle, this paper introduces Cyber Persistence Detector (\Sys), a novel system dedicated to detecting cyber persistence through provenance analytics. \Sys is founded on the insight that persistent operations typically manifest in two phases: the \quotes{persistence setup} and the subsequent \quotes{persistence execution}. By causally relating these phases, we enhance our ability to detect persistent threats. First, \Sys discerns setups signaling an impending persistent threat and then traces processes linked to remote connections to identify persistence execution activities. A key feature of our system is the introduction of \textit{pseudo-dependency edges} (\Edges), which effectively connect these disjoint phases using data provenance analysis, and \textit{expert-guided edges}, which enable faster tracing and reduced log size. These edges empower us to detect persistence threats accurately and efficiently. Moreover, we propose a novel alert triage algorithm that further reduces false positives associated with persistence threats. Evaluations conducted on well-known datasets demonstrate that our system reduces the average false positive rate by \UseMacro{fpr_avg_reduction}\% compared to state-of-the-art methods.

\end{abstract}


\section{Introduction}
\label{s:intro}

Advanced Persistent Threat (APT) attacks are increasingly leveraging Living-Off-the-Land Binaries (LOLBins), shifting the strategic focus from traditional malware to more nuanced persistence techniques. According to MITRE~\cite{mitre}, persistence techniques are defined as methods that adversaries use to keep access to systems across restarts, changed credentials, or other interruptions that could cut off their access. These techniques typically involve the installation of malicious software or the manipulation of legitimate scripts and tasks to ensure continuous, unauthorized remote access. Techniques such as reverse shells, SSH, Powershell Remoting, or other executables are often used for establishing and maintaining remote connections. Notably, persistence techniques were a key feature in nearly 75\% of cyberattacks in 2022~\cite{CrowdStrikeStudy2023}.

For instance, the Sandworm APT group used webshell persistence in multistage attacks~\cite{sandwormteam, sandwormcentreoncampaign, paswebshell}. The SolarWinds attack shows how persistence is crucial in APT campaigns, using scheduled tasks for this purpose~\cite{SUNBURST, SolarWindsAttack, SUNSPOT, mitre}. APT attackers often use a ``low and slow'' strategy, breaking their actions into phases with waiting periods to avoid detection. They gain initial access, establish persistence, disconnect to evade detection, and later reconnect for further malicious activities. This segmentation and strategic pausing are trademarks of the most stealthy APT attacks, as illustrated in Figure~\ref{fig:StealthinessByPersistence}.

In addressing the complexities of APT attacks, Provenance-based Intrusion Detection Systems (PIDS)~\cite{unicorn2020,protracer,hossain2017sleuth, winnower2018,holmes2019,nodoze2019,rapsheet2020,hossain2020combating,KAIROS,FLASH,shadewatcher,PROGRAPHER} have become essential by transforming audit logs into provenance graphs, providing causal relationships among system entities, such as processes and network sockets. In contrast, rule-based detection systems, such as Elastic~\cite{elasticdetectionrules}, Google Chronicle~\cite{chronicledetectionrules}, and Sigma~\cite{sigma}, which match audit logs against a predefined set of signatures, are industry standards for their capability to identify persistence threats.

Persistence techniques, as defined by MITRE~\cite{mitre}, often involve the misuse of \quotes{sensitive} system functionalities, such as Registry run keys~\cite{T1547001}. Current threat detection systems generate persistence attack alerts whenever these functionalities are accessed, irrespective of whether they are being misused by an attacker or legitimately used by a normal user. This approach fails to assess the consequences of the use of system functionalities, which might only become apparent later. Consequently, this leads to a high number of false positives; for example, a normal user adding entries in the Registry run keys or startup folder to launch programs upon log-on would trigger an alert, even though the action is benign. Conversely, if an attacker sets a Registry run key to initiate a command and control (C2) agent that connects back post-reboot, existing systems would identify the key's creation but might not link it to subsequent C2 activities due to the delay in their occurrence and the lack of a comprehensive context check necessary for accurate persistence detection.

\subsection{Limitations of Existing PIDS}

Anomaly-based or learning-based PIDS~\cite{provdetector2020,unicorn2020,shadewatcher,PROGRAPHER,wang2022threatrace,KAIROS,FLASH,MAGIC} aim to detect novel attacks with less prior knowledge. These PIDS model benign behavior from provenance graphs, and detect deviations from the modeled normal behavior. However, the assumption that attack-related activities are anomalous is not always true. Besides, they require representative training data which are not always available, and the training phase slows down the detection process. Last, learning-based PIDS are inherently more susceptible to concept drift and more vulnerable to evasion attacks~\cite{akul2023,Mukherjee2023}. Our evaluation in Section~\ref{s:eval} shows that state-of-the-art learning-based PIDS~\cite{KAIROS,FLASH} are not only slower but also less accurate in persistence detection, due to failure to learn persistence attacks' semantics.

\begin{figure}[!t]
  \begin{center}
    \includegraphics[width = 0.33\textwidth]{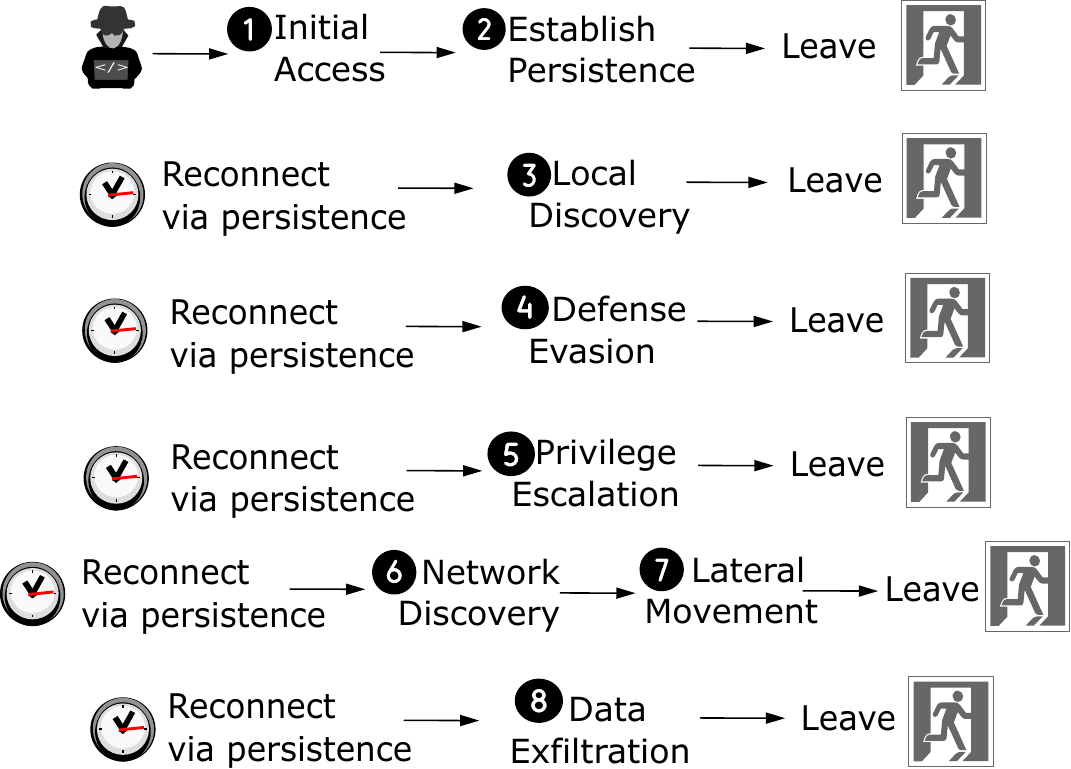}
  \caption[]{Stealthiness by persistence}
  \label{fig:StealthinessByPersistence}
  \end{center}
  \vspace{-4ex}
  \end{figure}

Existing heuristic-based PIDS also struggle with semantic understanding of persistence attacks, leading to significant challenges in connecting the dots across fragmented APT attack stages. This often results in incomplete and disconnected provenance (attack) graph reconstructions. Consider the attack scenario depicted in Figure~\ref{fig:StealthinessByPersistence}; heuristic-based PIDS, such as Holmes~\cite{holmes2019} and RapSheet~\cite{rapsheet2020}, often fail to piece together the full scope of an attack involving persistence. Instead, they produce isolated graphs, each representing only fragments of the APT attack. These systems triage attack graphs based on the number of APT stages, such as lateral movement and privilege escalation, contained within the graphs. If an attacker manages to fragment the attack graph into smaller, disconnected graphs, each segment inherently receives a lower severity score and is subsequently ranked lower for detection and investigation. Moreover, these heuristic-based PIDS require benign training data to quantify severity scores and filter false alarms.

\subsection{Limitations of Rule-based Persistence Detectors}
Current rule-based persistence detection systems, including popular solutions like Elastic~\cite{elasticdetectionrules} and Chronicle~\cite{chronicledetectionrules}, are plagued by a high rate of false positives (FPs). These systems typically analyze persistence techniques in isolation, neglecting the broader context of an attack. They often generate alerts for system activities that appear suspicious but are actually benign, leading to numerous false alarms. In an effort to mitigate these false positives, these systems may overly relax their detection rules. For instance, our evaluation in Section~\ref{s:eval} revealed that activities from programs in standard directories are automatically deemed benign without further scrutiny. This approach has inadvertently resulted in a significant increase in false negatives (FNs).

The narrow detection strategy has tangible consequences in Security Operations Centers (SOCs), where analysts spend roughly 30 minutes on each alert~\cite{alertatigue1,alertatigue2}, but with up to 99\% of these turning out to be FPs~\cite{Alahmadi2022}, leading to alert fatigue. To manage the deluge of alerts, thresholds are often set or high-volume alert rules are disabled, resulting in over two-thirds of alerts being disregarded~\cite{VectraStudy2023, alertatigue3}. This underscores the dire need for a system that can automatically reduce alert numbers.

\subsection{Our Approach and Contributions}

To address the issues with current persistence threat detection, we introduce Cyber Persistence Detector (\Sys), a novel system optimized for quick and accurate identification of persistence threats in enterprise networks. Our approach
\begin{itemize}
  \item[$\star$] avoids optimistic assumptions about persistence behavior,
  \item[$\star$] avoids the need for any training data,
  \item[$\star$] generates few FPs and FNs,
  \item[$\star$] triages persistence-related threat alerts, and
  \item[$\star$] generates accurate graphs for quick incident response.
\end{itemize}

\Sys is rooted in a thorough analysis of the MITRE ATT\&CK framework~\cite{mitrematrix} which is recognized as the most comprehensive and widely referenced directory of persistence threats.  We discovered that effective persistence attacks always consist of two phases: the persistence setup (e.g., creating a Registry run key) and the persistence execution (e.g., a remote connection initiated by that key). \textit{Persistence setup serves solely as preparation, whereas persistence execution exhibits attackers' true motives}. Recognizing this two-phase process is crucial, yet it is overlooked by existing detection systems.

Leveraging the above key insight, \Sys introduces a novel concept of \textit{pseudo-dependency edges} (\Edges) to connect persistence setup and persistence execution activities within system logs, creating a comprehensive provenance graph. \Sys starts by recording every persistence setup activity in system logs, and then specifically checks if there is a subsequent remote connection, i.e., potential persistence execution, that can be traced back to the persistence setup activity. If not, it does not raise an alert, significantly reducing false positives. If yes, it creates a \Edge, and then utilizes even more contextual indicators, detailed in Section~\ref{ss:FalsePositiveReduction}, before deciding if it is a persistence attack. The creation of a \Edge involves tracing processes in provenance graphs that initiate or receive remote connections, assessing their alignment with persistence execution and persistence setup as per our advanced detection rules. This approach enables \Sys to identify potential persistence activities through detection rules and to analyze processes engaged in remote connections, assessing their role in the broader scheme of a persistence threat. This dual-layered strategy, combining precise activity tracking with the creation of \Edges, empowers \Sys to surpass existing threat detectors.

Despite the advantages, relying solely on \Edges for enhancing persistence detection accuracy presents challenges. One major issue is that even the most comprehensive logging systems may not capture all connections, potentially leading to gaps in the provenance graph and, consequently, false negatives. Our research found that integrating Windows ALPC logs with system audit logs could bridge these gaps. However, this integration significantly increases log storage requirements. Instead of depending on ALPC logs, \Sys employs a novel technique of {\it expert-guided edges}, utilizing insights from process creation and system policy to refine the provenance graph and reduce log volume by an average of 37\%. While \Edges specifically aid in identifying persistence threats, expert-guided edges serve a broader purpose, improving overall efficiency in tracing and log management without being limited to detecting persistence threats.

Another challenge arises from some benign programs' behaviors that cause \Edges to be generated excessively, which can complicate the detection process. To counteract this, \Sys incorporates a sophisticated false positive reduction algorithm, which is informed by an in-depth analysis of APT behaviors. This ensures that only genuinely malicious activities are flagged. Our algorithm capitalizes on the crucial understanding that persistence is merely one component of a multi-stage APT kill chain, all of which need to be executed in concert to fulfill the attackers' objectives. By verifying the presence of related kill chain techniques and tactics within close proximity in the provenance graph, \Sys significantly enhances its capability to distinguish between benign and malicious actions, improving both the accuracy and reliability of threat detection.

Our system \Sys outperforms existing detection systems, as evidenced by evaluations on both public datasets and those derived from strictly implemented MITRE emulation plans~\cite{AdversaryEmulationLibrary}. It excels in reducing FPs by \UseMacro{fpr_avg_reduction}\% on average, effectively detecting true persistence attacks, and producing succinct attack graphs that explain the persistence setup and execution. The capability to pinpoint persistence setup and execution within a provenance graph and present it in context provides security analysts with actionable insights for further investigation, demonstrating \Sys's practicality. Mostly, \Sys has a response time of under a minute for its entire pipeline.

Unlike previous techniques, \Sys produces alarms satisfying all five properties of reliability, explainability, analytical depth, contextuality, and transferability as introduced in \cite{Alahmadi2022}. In its first stage, \Sys's detection rules avoid easily changeable indicators, such as hard-coded IP addresses and file hashes, which are common in rule-based security systems~\cite{elasticdetectionrules,sigma,chronicledetectionrules}, ensuring reliable detection. The attack graphs produced in the second stage of \Sys are both explainable and contextual, providing an analytical overview of the attack. Finally, the system's customizable weighting factors for indicators and the alert budget introduced in the third stage make \Sys highly transferable and adaptable for practical use.

The main contributions of our paper are:

\begin{itemize}[leftmargin=*,topsep=-1pt,itemsep=0pt]

\item We present a thorough analysis of cyber persistence attacks and propose \Sys, the first detection system specifically targeting persistence threats.

\item We introduce \textbf{pseudo-dependency edges} to causally relate disjoint persistence phases and improve the detection of these threats.

\item We propose the novel concept of \textbf{expert-guided edges} to enable efficient provenance tracing of persistence threats.

\item We identify critical insights on determining malicious persistence behaviors and propose \textbf{an alert triage algorithm} incorporating these insights to reduce false positives.

\item We implement and evaluate \Sys on diverse datasets, demonstrating better attack detection rates and provenance graph completeness versus state-of-the-art methods.

\end{itemize}

\begin{table*}
\centering
  \caption{Top 10 persistent (sub-)techniques}
  \resizebox{0.95\textwidth}{!}{%
  \begin{tabular}{ccccccccccc}
    \toprule
    \makecell{Registry Run Keys\\/ Startup Folder} & \makecell{Scheduled\\Task} & \makecell{Web\\Shell}  & \makecell{DLL\\Side-Loading} &        \makecell{External\\Remote Services} & \makecell{Windows\\Service} & \makecell{Domain\\Accounts} & \makecell{WMI Event\\Subscription} & \makecell{DLL Search\\Order Hijacking} & \makecell{Local\\Account} \\
    \midrule
    49 & 45 & 23 & 21 & 20 & 20 & 11 & 10 & 9 & 9 \\
    \bottomrule
  \end{tabular}}
  \label{tab:top10persistence}
\end{table*}
\begin{table*}
\centering
  \caption{Top 10 persistent APT groups}
  \resizebox{0.75\textwidth}{!}{%
  \begin{tabular}{ccccccccccc}
    \toprule
    \makecell{APT29} & APT41 & \makecell{Lazarus\\Group}  & \makecell{Sandworm\\Team} & \makecell{Kimsuky} & \makecell{APT28} & \makecell{APT39} & \makecell{APT3} & \makecell{Magic\\Hound} & \makecell{Threat\\Group-3390} \\
    \midrule
     25 & 16 & 12 & 11 & 10 & 10 & 8 & 8 & 8 & 8 \\
    \bottomrule
  \end{tabular}}
  \label{tab:top10APT}
\end{table*}

\section{Motivation}
\label{ss:motivation}

\subsection{APT Attack Stages} 
In a typical APT attack campaign, attackers gain \textit{initial access} to a victim organization mainly through program exploits or stolen credentials. Program vulnerabilities are often patched, and users frequently change passwords, making these methods unreliable for long-term operations. Thus, attackers \textit{establish persistence} using various methods for prolonged, reliable access. Post-persistence, they perform \textit{local discovery} to understand target systems, including security program details. To \textit{evade detection}, they select tools to bypass these security programs or deactivate them after \textit{privilege escalation}. Afterwards, attackers conduct \textit{network discovery} to find and \textit{move laterally} to other vulnerable machines. The hallmark of APT campaigns is not immediate impact but remaining undetected for long periods, aiming to \textit{collect} and \textit{exfiltrate data} or cause significant \textit{impact} at a strategic point. The ability to achieve persistence is critical for attackers' success.

\subsection{Persistence Prevalence} 
\label{ss:PersistencePrevalence}
We extracted data from MITRE ATT\&CK knowledge base~\cite{attackstixdata}, which includes adversary tactics and techniques based on real-world observations. Our statistical analysis on MITRE's database sheds light on the prevalence of persistence \mbox{(sub-)techniques} in the wild. There are 99 distinct (sub)-techniques in persistence tactic. We find out that 94 out of 136 APT groups (69\%) leveraged at least one persistence \mbox{(sub-)technique} in the past. We rank both persistence \mbox{(sub-)techniques} based on the number of APT groups that leveraged these techniques in the real world, and APT groups based on the number of persistence \mbox{(sub-)techniques} employed by them. Due to space limit, we only show the top 10 persistence (sub)-techniques in Table~\ref{tab:top10persistence}, and top 10 \quotes{persistent} APT groups in Table~\ref{tab:top10APT}.


\subsection{Why is Persistence Misunderstood in APT Detection?}
\label{ss:PersistenceMisconception}
Persistence is often seen in the wild, but ironically, despite its emphasis in the term A\underline{P}T, it is misunderstood in academic research. MITRE defines persistence as the ability to maintain access to victim systems across restarts, changed credentials, and other interruptions that could (temporally) cut off access. This can be achieved in two primary ways: either by attackers initiating a remote connection using stolen credentials (T1078.003) or placing a public key in the \texttt{SSH authorized\_keys} file (T1098.004); or by the attackers initiating a connection from within the victim system, such as by placing a new entry in the Registry run keys or startup folder (T1547.001), creating scheduled tasks or jobs (T1053), or inserting commands in Unix shell configuration files like \texttt{.bashrc} (T1546.004).  For further details regarding these techniques, we refer the reader to~\cite{mitrematrix}.

In our review of all public audit log datasets~\cite{OpTC,TransparentComputing,atlas} for evaluating PIDS, we identified a significant gap in the understanding of persistence. Effective persistence hinges on three key conditions: setting a trigger (like creating a Registry run key), linking code for remote connection to this trigger (e.g., placing an executable in the Registry run key), and the successful initiation of a remote connection when the trigger is activated. In practice, the third condition may not always be met, prompting APT actors, such as APT29 to deploy multiple persistence techniques to enhance their odds of maintaining presence in the target environment. Unfortunately, existing datasets often fulfill only the first condition, setting up a trigger, and frequently link an irrelevant value to it, thereby missing the second and third conditions. This oversimplification can be counterproductive, as normal programs often activate these persistence triggers, enabling attackers to blend into routine system activities. Unlike these datasets, each of MITRE's eleven emulation plans~\cite{AdversaryEmulationLibrary}, based on real-world APT behaviors, incorporates at least two persistence techniques without such oversimplification. Thus, we utilize these emulation plans to evaluate \Sys as detailed in Section~\ref{s:eval}.

\begin{figure*}[!t]
    \centering
    \includegraphics[width=0.99\linewidth]{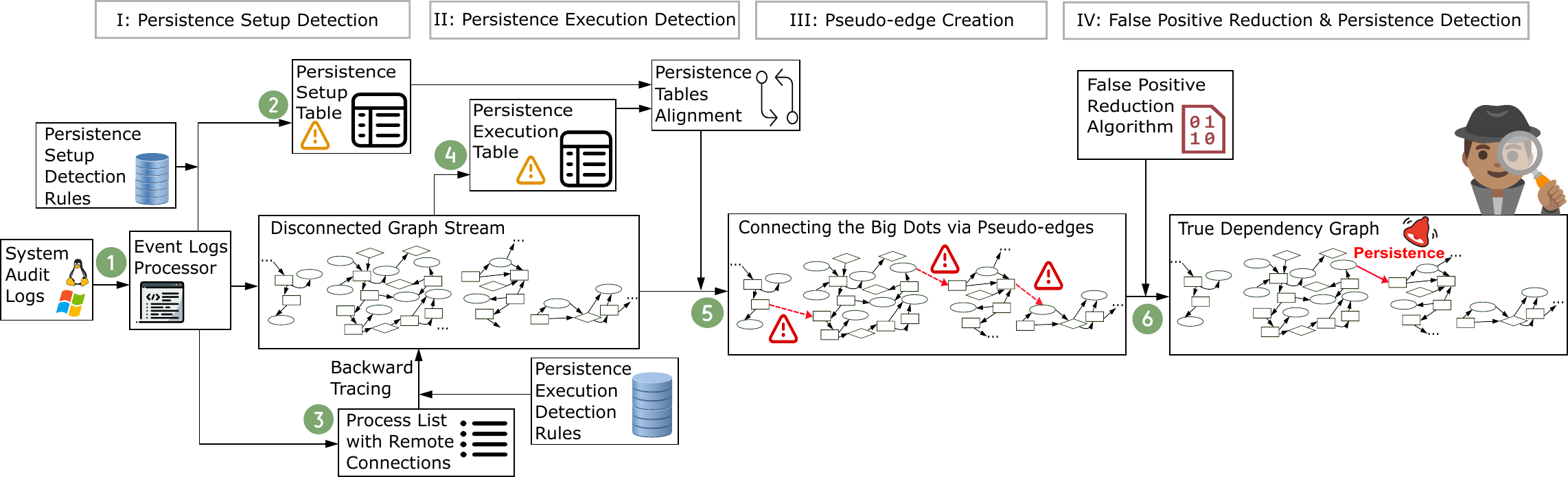}
    \caption[]{\Sys overview. \Sys implements a four-step approach for detecting persistence threats, starting with the creation of a persistence setup table from audit logs that tracks potential setup actions. It then traces processes with remote connections to form sub-graphs, which are evaluated against execution rules and aligned with setup actions to form atomic graphs linked by a \Edge. The process is refined through the introduction of \Edge strength and a false positive reduction algorithm.}
    \label{fig:SystemOverview}
    \vspace{-2ex}
  \end{figure*}
  
\section{Threat Model}
\label{s:ThreatModel}
Our system, like other PIDS \cite{unicorn2020,protracer,hossain2017sleuth, winnower2018,holmes2019,nodoze2019,rapsheet2020,hossain2020combating}, considers firmware, OS, and our logging systems in trusted computing base (TCB). Unlike \cite{holmes2019,nodoze2019,rapsheet2020,unicorn2020}, we do not presume attacks end before an OS reboot. Our system is unique in linking provenance attack graphs across reboots using \textit{\Edges}. Our logging, as a Windows service or a \texttt{Systemd} service on Linux, restarts post-reboot, but an OS reboot may miss system audit events. Notably, attack-related processes starting before our logging service are not logged, often indicating persistence attacks. On Windows, persistence can be achieved through \quotes{Create or Modify System Process: Windows Service} (T1543.003), and on Linux through \quotes{Create or Modify System Process: Systemd Service} (T1543.002).  We can trace back to the root process in logs despite missing initial process creation events, using parent process GUID/ID. Our tools System Monitor~\cite{sysmon} on Windows and Auditd~\cite{linuxaudit} on Linux record these IDs.

\PP{Technique Coverage} Of those 99 distinct \mbox{(sub-)techniques} in persistence tactic, only half of them were knowingly used by at least one APT group in the past according to MITRE's database. We only created detection rules for about a third of all persistence \mbox{(sub-)techniques}, which all fall into the \quotes{significant} half of persistence \mbox{(sub-)techniques} and include all top 10 persistence \mbox{(sub-)techniques}. We have excluded all persistence \mbox{(sub-)techniques} related to macOS, cloud infrastructures, pre-OS boot like Bootkit.

\section{System Design}
\label{s:design}

\Sys employs a four-step approach for accurate persistence detection, as illustrated in Figure~\ref{fig:SystemOverview}. Initially, \Sys processes system event logs to create a \textit{persistence setup table}, recording activities that match persistence setup detection rules, such as new account creations or Registry run key additions. Next, it identifies processes with remote connections in the event stream, performing individual backward tracing to generate sub-graphs for each. These sub-graphs are then checked against our persistence execution detection rules, with matches recorded in a \textit{persistence execution table}. In the third step, \Sys aligns entries from persistence execution table with persistence setup table based on TTP labels\footnote{TTP stands for Tactics, Techniques, and Procedures. An exemplary TTP label is T1547.001.}, temporal order, and specific attributes. It creates a \textit{persistence setup atomic graph}, i.e., a minimal sub-graph directly related to persistence, and a \textit{persistence execution atomic graph} when an alignment is found, and then links them with a \Edge. For evaluation purposes (Section~\ref{s:eval}), steps 2 and 3 are combined as step 2 alone does not yield direct alerts. The final step introduces \textit{\Edge strength} and a false positive reduction algorithm to ensure only significant events are connected.

\subsection{Persistence Threat Detection}
Our persistence setup detection rules are created by studying persistence \mbox{(sub-)techniques} described in MITRE ATT\&CK Matrix, hundreds of persistence-related threat reports, and red team tools that provide visibility into low-level code related to persistence attacks, like Atomic Red Team~\cite{atomicredteam}. We then improved our persistence setup detection rules by studying and incorporating open-source detection rules from popular rule repositories like Sigma~\cite{sigma} and Elastic~\cite{elasticdetectionrules}. However, as discussed in Section~\ref{ss:OpenSourceDetectionRules} in details, our rules are different from the ones in those repositories. We will commit our persistence detection rules to these open-source rule repositories for the public's benefit.\footnote{The first author of this article has been submitting accepted commits for non-persistence-related TTP detection rules to those repositories.}
The persistence setup detection process is mostly a straightforward string match process against information inside a single system log event.
But some rules have a few more rule conditions, which require information across several log events. For this, we use \quotes{sequenced} query of EQL, a query language specifically designed for threat hunting~\cite{eql}.  Indicative strings include especially file paths, Registry locations, process names and command lines etc. Note that these strings are characteristic to the persistence \mbox{(sub-)techniques}. For instance, to implement the persistence sub-technique T1547.001 (Registry run keys), one of a few known Registry locations \textit{must} be modified.

\begin{figure*}[!t]
  \centering
  \includegraphics[width=0.88\linewidth]{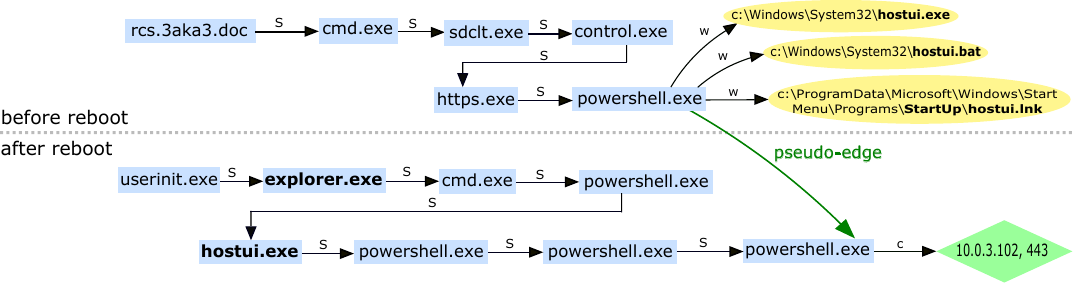}
  \caption{
    A persistence attack graph automatically generated by \Sys on the EP-APT29-1 dataset. It uses rectangles for processes, ovals for files / Registry keys, and diamonds for network sockets. Annotations include S=Start, W=Write, C=Connect. The graph successfully pinpoints T1547.001 (Boot or Logon Autostart Execution: Registry Run Keys / Startup Folder). The upper section reveals persistence setup: a malicious Microsoft Word-like program (\texttt{.doc}) starts, resulting in a Powershell instance and a shortcut creation in the Windows startup folder. This shortcut leads to another dropped malicious program, \texttt{hostui.exe}. The lower section, post-reboot, shows persistence execution: \texttt{explorer.exe} auto-executes startup folder shortcuts, triggering malicious Powershell code and connecting to the attacker. Indicative strings are bolded for clarity. \Sys forms a \Edge linking the process initiating persistence setup with the one managing the remote connection, i.e., the c2 agent.}
    \label{fig:APT29Persistence}
\end{figure*}

Simply matching system events against those detection rules generates a huge volume of alerts in practice. To reduce false alerts, \Sys first spots every process initiating or accepting remote connection(s) in the event stream, then performs backward tracing on these processes individually. During the backward tracing, \Sys inspects whether its provenance graph contains system activities matching our persistence execution detection rules. Like persistence setup detection rules, our persistence execution detection rules contain other indicative strings incorporating file paths, Registry locations, process names etc. These strings are also characteristic to the persistence \mbox{(sub-)techniques}. For example, to \quotes{activate} the persistence sub-technique T1547.001 (Registry run keys), one of a few known Registry locations \textit{must} be read by exactly the process \texttt{explorer.exe}, and it \textit{must} rely on this process to (directly or indirectly) start the malicious process.

Table~\ref{tab:detectionRulesEffectiveness} demonstrates how sensitive these detection rules are. Note that we assume the integrity of the operating system including its native built-in system programs. We stress that, at its stage 1, \Sys sacrifices specificity for sensitivity in its detection results, in order to not miss a single potential persistence attack. In other words, \Sys will not have false negatives at this stage, but at the cost of having many false positives. The optimization techniques introduced in \Sys and discussed in Section~\ref{ss:OpenSourceDetectionRules} only improve its stage 1's specificity, and do not impact the sensitivity. Besides, we argue that the evasion techniques for SIEM (Security Information and Event Management) rules introduced in \cite{Uetz2024} have only limited impact on our detection rules. Because, as discussed in Table~\ref{tab:detectionRulesEffectiveness}, we mostly do not rely on recorded command lines or code executed by attackers. Rather, we use indicative file paths, Registry locations and system process names. These strings are immutable under the assumption of OS integrity.   

\begin{table*}
  \caption{Detection rule sensitivity for the top 10 persistent techniques. TPR = True Positive Rate, \V = Yes, \Vc = Almost.}
  \resizebox{0.94\textwidth}{!}{%

  \begin{tabularx}{\textwidth}{ccX}
    \toprule
    (Sub-)techniques & \makecell{Detection rules\\with TPR=1?} & Remark\\
    \midrule
     \makecell[t]{Registry Run\\ Keys / Startup\\ Folder} & \V & During persistence setup, new entries \textit{must} be added to the the standard Registry run keys locations or standard Startup folders. During persistence execution, the corresponding new entries \textit{must} be read by the system process \texttt{explorer.exe}. The malicious process \textit{must} be ultimately started by \texttt{explorer.exe}.\\
     \midrule
     \makecell[t]{Scheduled\\Task} & \V  & During persistence setup, one of a few Windows task creation programs / Powershell \texttt{cmdlets} / API \textit{must} be called, and file modification in the Windows standard Tasks folder \textit{must} be undertaken. During persistence execution, the malicious process \textit{must} be ultimately started by the system process \texttt{svchost.exe} with the flags \quotes{\texttt{-k netsvcs -p -s Schedule}}.\\
    \midrule
     \makecell{Web Shell} & \Vc & During persistence setup, a file containing executable code like PHP is \textit{very likely} dropped to the web root directory like \texttt{/var/www/html}. During persistence execution, the malicious process \textit{must} be ultimately started by the web server program like \texttt{apache2}.\\
     \midrule
     \makecell[t]{DLL\\Side-Loading} & \V  & During persistence setup, a DLL file \textit{must} be dropped to the file system. During persistence execution, that DLL file \textit{must} be loaded by a process initiating or accepting remote connection(s).\\
     \midrule
     \makecell[t]{External\\Remote Services} & \V  & During persistence setup, a common remote access program \textit{must} be installed, and its executable file \textit{must} be dropped to the file system. During persistence execution, that program \textit{must} initiate or accept remote connection(s). \\
    \midrule
     \makecell[t]{Windows\\Service} & \V  & During persistence setup, a new entry \textit{must} be added to the standard Registry location for Windows services. During persistence execution, the corresponding new entry \textit{must} be read by the system process \texttt{services.exe}. The malicious process \textit{must} be ultimately started by \texttt{services.exe}.\\
    \midrule
     \makecell[t]{Domain\\Accounts} & \V  & During persistence setup, a new entry \textit{must} be created in the domain controller's standard Active Directory database stored in the file system. During persistence execution, this new account \textit{must} be used for logging into target systems. \\
     \midrule
    \makecell[t]{WMI Event\\Subscription} & \V  & During persistence setup, one of a few Windows WMI event creation programs / Powershell \texttt{cmdlets} / API \textit{must} be called, and file modification in the Windows standard WMI event repository \textit{must} be undertaken. During persistence execution, the malicious process \textit{must} be ultimately started by the system process \texttt{wmiprvse.exe}. \\
    \midrule
    \makecell[t]{DLL Search\\Order Hijacking} & \V  & During persistence setup, a DLL file \textit{must} be dropped to the file system. During persistence execution, that DLL file \textit{must} be loaded by a process initiating or accepting remote connection(s). \\
     \midrule
    \makecell[t]{Local\\Account} & \V  & During persistence setup, a new entry \textit{must} be created in the standard user information database stored in the local file system. During persistence execution, this new account \textit{must} be used for logging into target systems. \\
    \bottomrule
  \end{tabularx}}
  \label{tab:detectionRulesEffectiveness}
\end{table*}

\begin{algorithm}[!t]
  \scriptsize
  \DontPrintSemicolon
  \SetKwInOut{Input}{Inputs}
  \SetKwInOut{Output}{Output}
  
  \SetKwFunction{CreatePE}{\sc{CreatePseudoEdge}}
  \SetKwFunction{GetPS}{\sc{GetPersistenceSetup}}
  \SetKwFunction{GetPE}{\sc{GetPersistenceExecution}}
   
  \Fn{\CreatePE{Events \bigEI{}}}{%
  \tcc{Get a list of persistent setup events}
  $L_{{<\mathcal{E{\alpha}},\mathcal{L{\alpha}},\mathcal{T{\alpha}}>}}$ $\leftarrow$ \textsc{GetPersistenceSetup}(\bigEI{}) 
  
   \tcc{Get a list of processes with remote conn.}
  $L_{{<\mathcal{P{\gamma}}>}}$ $\leftarrow$ \textsc{GetProcessWithRemoteConnection}(\bigEI{}) 
  
  \ForEach{\bigPI{\gamma} $\in$ $L_{<\mathcal{P}>}$}{     \label{line:CheckProcess1}
    
    \tcc{Get a list of persistent exec. events}
    $L_{{<\mathcal{E{\gamma}},\mathcal{L{\gamma}},\mathcal{T{\gamma}}>}}$ $\leftarrow$ \textsc{GetPersistenceExecution}(\bigPI{\gamma})
    
  	\ForEach{(\bigEI{\gamma},\bigLI{\gamma},\bigTI{\gamma}) $\in$ $L_{{<\mathcal{E{\gamma}},\mathcal{L{\gamma}},\mathcal{T{\gamma}}>}}$}{
     
     \ForEach{(\bigEI{\alpha},\bigLI{\alpha},\bigTI{\alpha}) $\in$ $L_{{<\mathcal{E{\alpha}},\mathcal{L{\alpha}},\mathcal{T{\alpha}}>}}$}{
     
     \If{\bigLI{\gamma} $==$ \bigLI{\alpha}}{
     
       \If{\bigTI{\gamma} $>$  \bigTI{\alpha}}{    \label{line:CheckProcess2}
       
       $AG_{\gamma}$ $\leftarrow$ \textsc{GetAtomicGraph}(\bigEI{\gamma})  \label{line:GetAtomicGraph}
       
       $AG_{\alpha}$ $\leftarrow$ \textsc{GetAtomicGraph}(\bigEI{\alpha}) 
       
       \tcc{Create a \Edge}
       $PAG{(\gamma,\alpha)}$  $\leftarrow$ $AG_{\gamma}$ $\cup$ $AG_{\alpha}$ \label{line:CreateEdge}
       
       $L_{<PE,PAG>}$  $\leftarrow$ $L_{<PE,PAG>}$ $\cup$ $<PE(\gamma,\alpha),PAG(\gamma,\alpha)>$ 
       }
      }
     }
    }
   }
   \Return $L_{<PE,PAG>}$
  }

  \BlankLine
  \BlankLine

  \Fn{\GetPS{\bigEI{}}}{%
    \ForEach{$Rule$ $\in$ $L_{PersistenceSetupRule}$}{
       \ForAll {$Condition$ $\in$ $Rule$}{
          $satisfied$  $\leftarrow$ \textsc{CheckCondition}($Condition$, \bigEI{})
             
         \If{$satisfied$}{
          $L_{{<\mathcal{E{\alpha}},\mathcal{L{\alpha}},\mathcal{T{\alpha}}>}}$ $\leftarrow$  $L_{{<\mathcal{E{\alpha}},\mathcal{L{\alpha}},\mathcal{T{\alpha}}>}}$ $\cup$ (\bigEI{},\bigLI{},\bigTI{})
     } 
    }
   }
   \Return $L_{{<\mathcal{E{\alpha}},\mathcal{L{\alpha}},\mathcal{T{\alpha}}>}}$
  }
  
  \BlankLine
  \BlankLine
  
  \Fn{\GetPE{\bigPI{\gamma}}}{%
  
   $L_{{<\mathcal{E{\kappa}}>}}$ $\leftarrow$ \textsc{TraversalBackward}(\bigPI{\gamma}, \bigEI{}) \label{line:TraversalBackward}
    
   \ForEach{\bigEI{\kappa} $\in$ $L_{{<\mathcal{E{\kappa}}>}}$}{    
    
    \ForEach{$Rule$ $\in$ $L_{PersistenceExecutionRule}$}{
    
      \ForAll {$Condition$ $\in$ $Rule$}{
      
       $satisfied$  $\leftarrow$ \textsc{CheckCondition}($Condition$, \bigEI{\kappa})
             
       \If{$satisfied$}{
        $L_{{<\mathcal{E{\alpha}},\mathcal{L{\alpha}},\mathcal{T{\alpha}}>}}$ $\leftarrow$  $L_{{<\mathcal{E{\alpha}},\mathcal{L{\alpha}},\mathcal{T{\alpha}}>}}$ $\cup$ (\bigEI{\kappa},\bigLI{\kappa},\bigTI{\kappa})
      } 
     }
    }
   }
   \Return $L_{{<\mathcal{E{\gamma}},\mathcal{L{\gamma}},\mathcal{T{\gamma}}>}}$
  }
  
  \caption{\sc{pseudo-edge Creation}}
  \label{alg:pseudo-edge_creation}
\end{algorithm}

For every detected potential malicious process (with remote connections) from above, \Sys checks if it has corresponding entries in the persistence setup table, based on the TTP labels, some TTP-specific attributes, and happens-before relationship (Algorithm~\ref{alg:pseudo-edge_creation} Lines~\ref{line:CheckProcess1}-\ref{line:CheckProcess2}). If an alignment is found, a persistence execution atomic graph is created (Algorithm~\ref{alg:pseudo-edge_creation} Line~\ref{line:GetAtomicGraph}), which includes only information related to persistence execution. Likewise, a persistence setup atomic graph containing only critical attack information is also generated. Then \Sys creates a \Edge to connect the persistence setup atomic graph and the persistence execution atomic graph (Algorithm~\ref{alg:pseudo-edge_creation} Line~\ref{line:CreateEdge}), resulting into a succinct and insightful persistence attack graph, as shown in Figure~\ref{fig:APT29Persistence}.

\subsection{Expert-guided Edges}

In our experiments, we found limitations in linking system entities solely based on Windows' Process Monitor / System Monitor logs. Specifically, during T1543.003 (Create or Modify System Process: Windows Service) persistence setup, attackers often use \texttt{sc.exe} to create a malicious Windows service. This results in a new Registry key under \texttt{HKLM\textbackslash SYSTEM\textbackslash CurrentControlSet\textbackslash Services}, the basis for our detection rule. This approach, focusing on an immutable Registry location, is more reliable than relying on command lines, which are easily bypassed, as recent research shows~\cite{Uetz2024}.

In the logs, the corresponding Registry key modifications appear to be done by \texttt{services.exe}, not \texttt{sc.exe}, with no apparent link between the two. Further research and consultation with the Windows Developer Reference~\cite{Allievi2022} revealed that the communication between these processes occurs through ALPC, a Windows inter-process communication (IPC) method not typically logged by standard frameworks. Additional logging via Windows ETW \quotes{NT Kernel Logger} confirmed the link but resulted in excessively large datasets due to ALPC's widespread use.

\begin{figure}[!t]
  \centering
  \includegraphics[width=0.98\linewidth]{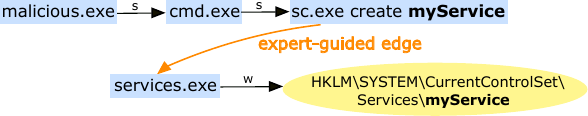}
  \caption{An expert-guided edge is created during reconstruction of a T1543.003 persistence setup attack graph. An attacker-controlled malicious process leverages LOLBins to create a malicious service for persistence. The indicative Registry key is however modified by a Windows system process, to which no link from the malicious process can be built using logs from standard logging frameworks.}
  \label{fig:expertEdge}
\end{figure}

To address this, we introduce \textit{expert-guided edges} in \Sys. These edges are formed by applying specialized parsing rules during log processing for provenance graph generation. This method embeds expert knowledge about process creation routines and operating system policies into the backward and forward tracing process. For example, we can link \texttt{sc.exe} and \texttt{services.exe} if \texttt{services.exe} modifies a Registry key under the specified location right after \texttt{sc.exe} is executed with the same service name, as illustrated in Figure~\ref{fig:expertEdge}. This approach offers three benefits: faster search process, reduced dependency explosion, and bridging gaps otherwise impossible to close.

Similarly, we observe missing links on Linux using Auditd logs even though we are monitoring a very extensive list of syscalls. Specifically, during T1543.002 (Systemd Service) persistence setup, an indicative file \texttt{/etc/systemd/system/*.service} is created. However, during persistence execution, we cannot observe the same file being accessed, but rather a closely related in-memory file \texttt{/sys/fs/cgroup/system.slice/*.service/*} with the same service file name. An edge cannot be built between the corresponding sub-graphs if using the traditional \quotes{write and read on the same file node} principle. Hence we use an expert-guided edge to resolve this issue. Algorithm~\ref{alg:expert_guided_edge_creation} describes the creation of expert-guided edges, and is applied in Line~\ref{line:TraversalBackward} of Algorithm~\ref{alg:pseudo-edge_creation}. Note that we create expert-guided edges under the assumption that the integrity of the OS itself is not compromised. We show the log storage reduction rate by introducing expert-guided edges in Section~\ref{ss:logReduction}.

\begin{algorithm}[!t]
  \scriptsize
  \DontPrintSemicolon
  \SetKwInOut{Input}{Inputs}
  \SetKwInOut{Output}{Output}
  \Input{System audit log events \bigEI{}; \\
    List of critical system processes $L_{<\mathcal{P}>}$}
  \Output{List of dependency path $L_{<P>}$}
  \BlankLine
  
  \ForEach{\bigEI{\kappa} $\in$ \bigEI{}}{ 
    \tcc{Get the process of current event}
    \bigPI{\kappa} $\leftarrow$ \textsc{GetSubject}(\bigEI{\kappa}) 

      \If{\bigPI{\kappa} $\in$ $L_{<\mathcal{P}>}$}{
        \tcc{Add dependency path to the standard routine nodes}
        $P$ $\leftarrow$ \textsc{AddPathToRoutineNodes}(\bigPI{\kappa})
        
        $L_{<P>}$ $\leftarrow$ $L_{<P>}$ $\cup$ $P$
      } 
   }

  \Return $L_{P}$
  \caption{\sc{expert-guided edge creation}}
  \label{alg:expert_guided_edge_creation}
\end{algorithm}

\subsection{False Positive Reduction} 
\label{ss:FalsePositiveReduction} 

\begin{figure*}
  \centering
  \includegraphics[width=0.78\linewidth]{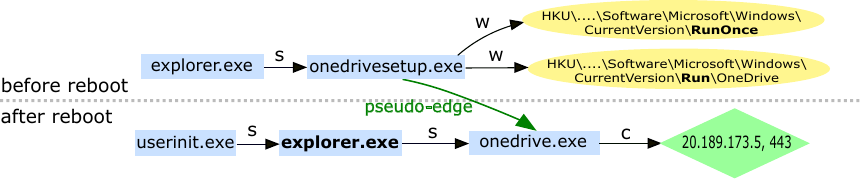}
  \caption{A false-positive persistence attack graph automatically generated by \Sys on the EP-APT29-1 dataset. This graph wrongly classifies an instance of T1547.001. It turns out to be a benign program, i.e., Microsoft OneDrive, leveraging Registry run keys for updates. It in fact connects back to an IP address belonging to Microsoft Corporation.}
  \label{fig:FalsePositive}
\end{figure*}

Creating a \Edge that accurately indicates a persistence attack is challenging. 
This is mainly due to the fact that many benign programs use Registry run keys (T1547.001), Windows services (T1543.003), scheduled tasks (T1053.005) etc. for some program-specific routines involving remote connections, leading to an inadequate amount of false-positive persistence attack graphs. For instance, some benign programs create a Windows service to check and download updates periodically, like Adobe Acrobat's Update Service (\texttt{armsvc.exe}), mimicking persistence behaviors and triggering false-positive \Edges in \Sys. While the Windows service creation itself mimics persistence setup, the resulted update downloads (involving remote connections) later on mimic persistence execution. Similarly, programs such as Google Chrome and Microsoft OneDrive use Registry run keys for updates, frequently leading to false alarms. Figure~\ref{fig:FalsePositive} shows a typical false-positive persistence attack graph generated by \Sys after its stage 2. 

To address this, we developed a false positive reduction algorithm using contextual indicators to differentiate between benign and malicious activities. This algorithm introduces \textit{\Edge strength} and ranks \Edges based on a calculated threat score. We categorize \Edges as either causality-based or correlation-based for improved detection accuracy, due to the difference in the nature of various persistence techniques. Correlation-based \Edges, related to login account techniques like T1098, T1136, and T1078, are less reliable due to the uncertainty of user identity behind consecutive logins. These receive a `penalty' weight (less than 1) in the anomaly score assignment (Equation~\ref{eq:SumScore}). For precise detection, we utilize context not only from the cyber-kill-chain tactic and technique levels, but also from the program execution level.

\subsubsection{Causality-based \Edges} 

We formulate the following indicators based on studying APT behaviors in real-world attacks.

\begin{itemize}[leftmargin=*,topsep=-1pt,itemsep=0pt]
\item \textbf{Degree of indirection} in both persistence setup and persistence execution. As observed in APT29’s real-world behaviors, multiple indirection is implemented to start the command and control (c2) agent program. That is, when the malicious shortcut file in a Windows startup folder is read, the \texttt{explorer.exe} process runs a normal-looking batch file linked to the shortcut file (indirection 1). The resulted \texttt{cmd.exe} process starts a \texttt{powershell.exe} process as instructed in the batch file (indirection 2), which in turn starts another normal-looking process (indirection 3). This process then starts another \texttt{powershell.exe} process (indirection 4), which again starts another \texttt{powershell.exe} process (indirection 5) before contacting the c2 server. This obviously deviates from normal programs' use of Windows startup folders. 

\item \textbf{Credential access tactic}, as observed in APT actors' past behaviors, e.g., APT29~\cite{apt29}, Sandworm~\cite{sandwormteam}, Wizard Spider~\cite{wizardspider}, is almost always executed as an attempt to obtain the \quotes{low-hanging fruits} persistence. 

\item \textbf{Persistence techniques are often executed together}, as they together will likely contribute to more reliable persistence. 

\item APT actors tend to execute persistence techniques \textbf{right after initial access or lateral movement}. For example, Wizard Spider achieved persistence right after initial compromise on the first victim machine, and then persistence on the second victim machine right after lateral movement.

\item APT actors tend to execute some \textbf{discovery techniques before} persistence techniques. However, we find that this indicator tends to be less reliable and more \quotes{noisy} than other indicators. To counter this, we assign the smallest weighting factor to this indicator during anomaly score calculation in Equation~\ref{eq:SumScore}.  
\end{itemize}

\subsubsection{Correlation-based \Edges} 
We further classify correlation-based \Edges into two types. 

\paragraph{Type 1 - persistent initial re-access}
We identified the following indicators for this type of persistence.
\begin{itemize}[leftmargin=*,topsep=-1pt,itemsep=0pt]
\item Observation of \textbf{credential access tactic} is also an indicator for correlation-based \Edges. For instance, OS credential dumping (T1003) is often performed on victim computers, and the stolen credentials are used for the remote re-connection. Besides, the \textbf{usage intensity}, i.e., occurrence of the same technique, and \textbf{extensiveness}, i.e., variation of attempted credential access techniques, can be a weighting factor. For example, during one of Sandworm's engagements, they uploaded an executable to its target machine for dumping web credentials, and another executable for key-logging a valid user RDP session to obtain domain credentials.

\item The accessing computer does not have a \textbf{legit FQDN} or \textbf{computer account} in the domain. This is more likely a malicious persistent initial re-access, as attackers typically do not physically own a domain joined computer. 

\item \textbf{Local account creation} after lateral movement, e.g., in a WinRM~\cite{winrm} session.

\item \textbf{Installing, activating and enabling standard remote access tools} like VNC and RDP server. For instance, Carbank~\cite{carbanak} installed a VNC server on its victim machine for persistence after capturing credentials, and opened the corresponding port on firewall. 
\end{itemize}

\paragraph{Type 2 - persistent lateral movement}
Type 2 correlation-based \Edges are a special kind of \Edges, which represent an intersection between persistence tactic and lateral movement tactic. The following indicators are extracted from analyzing real-world APT attacks.
\begin{itemize}[leftmargin=*,topsep=-1pt,itemsep=0pt]
\item \textbf{Remote system discovery} (T1018) is performed on one domain-joined computer before a remote connection is initiated from this computer to another computer in the network. For instance, APT29 used LDAP queries to enumerate other hosts in the domain before creating a remote Powershell session to a secondary victim.

\item \textbf{Ingress tool transfer} (T1105) is performed from one domain-joined computer to another computer before the remote connection.

\item \textbf{Credential access tactic} is performed on one domain-joined computer before a remote connection. For instance, Sandworm has stolen SSH keys on the first compromised computer and then use these credentials to move laterally to a second computer. 

\item \textbf{Local account creation} after or during lateral movement.

\item \textbf{Installing, activating and enabling standard remote access tools} like VNC and RDP server.
\end{itemize}

Taking into account above indicators, \Sys calculates an anomaly/threat score, and uses this score to quantify \Edge strength and rank \Edges. By doing so, we ensure that the most likely malicious \Edges are always investigated first by a security analyst. During the final stage of \Sys, as formulated in Algorithm~\ref{alg:fpReduction}, a \Edge is first classified into one of the three categories above (Line~\ref{line:GetCategory}). Then the corresponding persistence attack graph is automatically 1) analyzed on to extract features related to indicators from above, e.g., the degree of indirection during both persistence setup and persistence execution, and 2) further explored on to include more contextual information and search for the existence of other indicators from above, i.e., related attack steps in a cyber-kill-chain, e.g., whether credential access is observed in its dependency graph. The dependency graph expands on the persistence attack graph, and is therefore more verbose. That is, our false positive reduction algorithm takes as input all indicators found in the succinct persistence attack graph as well as in its more verbose dependency graph. In the following, we explain how these indicators are adopted for anomaly score assignment in three equations.

\begin{algorithm}
  \footnotesize
  \DontPrintSemicolon
  \SetKwInOut{Input}{Inputs}
  \SetKwInOut{Output}{Output}
  \Input{System audit log events \bigEI{}; \\
    List $L_{<PE,PAG>}$ of \Edge and persistence attack graph pairs; \\
    List $L_{<ind-PAG>}$ of indicators inside persistence attack graphs; \\
    List $L_{<ind-DG>}$ of indicators inside dependency graphs; \\
    Max persistence-edge alert number \bigNI{}} 
  \Output{List $L_{<PE,AS>}$ of persistence edge and its anomaly score pairs} 
  \BlankLine
  \ForEach{ $<PE(\gamma,\alpha),PAG(\gamma,\alpha)>$ $\in$ $L_{<PE,PAG>}$}{ 
    $AS_{PE(\gamma,\alpha)}$ $\leftarrow$ $0$
    
    $L_{<AS_{PE}>}$ $\leftarrow$ $0$
  
    \tcc{Classify \Edge}
    $PE'(\gamma,\alpha)$ $\leftarrow$ \textsc{GetCategory}($PE(\gamma,\alpha)$) \label{line:GetCategory}
    
    \tcc{Select indicators based on \Edge type}
    $L'_{<ind-PAG>}$ $\leftarrow$ \textsc{GetIndicators}($PE'(\gamma,\alpha)$, $L_{<ind-PAG>}$) \label{line:GetIndicators1}

    $L'_{<ind-DG>}$ $\leftarrow$ \textsc{GetIndicators}($PE'(\gamma,\alpha)$, $L_{<ind-DG>}$) \label{line:GetIndicators2}
     
    \ForEach{ $indicator$ $\in$ $L'_{<ind-PAG>}$}{ \label{line:CalculateScore1Start}
       $AS_{indicator}$ $\leftarrow$ \textsc{CalculateScore1}($indicator$, $PAG(\gamma,\alpha)$)  \label{line:CalculateScore1}
       
       $L_{<AS_{PE}>}$ $\leftarrow$ $L_{<AS_{PE}>}$ $\cup$ $AS_{indicator}$
      } \label{line:CalculateScore1End}
 
    ($L_{{<\mathcal{E{\delta}}>}}$, $DG(\delta)$) $\leftarrow$ \textsc{TraversalBackward}($PAG(\gamma,\alpha)$, \bigEI{}) \label{line:TraversalBackwardFPR}
    
    ($L_{{<\mathcal{E{\eta}}>}}$, $DG(\eta)$) $\leftarrow$ \textsc{TraversalForward}($PAG(\gamma,\alpha)$, \bigEI{})  \label{line:TraversalForwardFPR}
    
    $DG(\kappa)$ $\leftarrow$ \textsc{MergeGraph}($DG(\delta)$, $DG(\eta)$)  \label{line:MergeGraph}
           
    $L_{{<\mathcal{E{\kappa}}>}}$ $\leftarrow$ $L_{{<\mathcal{E{\delta}}>}}$ $\cup$ $L_{{<\mathcal{E{\eta}}>}}$ 
    
    \ForEach{\bigEI{\kappa} $\in$ $L_{{<\mathcal{E{\kappa}}>}}$}{    \label{line:CalculateScore2Start} 
           
      \ForEach{ $indicator$ $\in$ $L'_{<ind-DG>}$}{ 
      
       $satisfied$ $\leftarrow$ \textsc{CheckIndicator}($indicator$, \bigEI{\kappa})
          
       \If{$satisfied$}{
       $AS_{indicator}$ $\leftarrow$ \textsc{CalculateScore2}($indicator$, $DG(\kappa)$) \label{line:CalculateScore2}
       
       $L_{<AS_{PE}>}$ $\leftarrow$ $L_{<AS_{PE}>}$ $\cup$ $AS_{indicator}$
       }
      }
     } \label{line:CalculateScore2End} 
     
     $AS_{PE(\gamma,\alpha)}$ $\leftarrow$ \textsc{SumScore}($L_{<AS_{PE}>}$) \label{line:SumScore}
     
     $L_{<PE,AS>}$ $\leftarrow$ $L_{<PE,AS>}$ $\cup$ $AS_{PE(\gamma,\alpha)}$ \label{line:AddToList}
    }
    
    $L_{<PE,AS>}$ $\leftarrow$  \textsc{SortByScore}($L_{<PE,AS>}$) \label{line:SortByScore}
    
    $L_{<PE,AS>}$ $\leftarrow$  \textsc{RemoveByBudget}($L_{<PE,AS>}$, \bigNI{})  \label{line:RemoveByBudget}
    
  \BlankLine
  \Return $L_{<PE,AS>}$
  \caption{\sc{false positive reduction}}
  \label{alg:fpReduction}
\end{algorithm}

First, we calculate the anomaly score of an indicator observable from the persistence attack graph as follows (Line~\ref{line:CalculateScore1} in Algorithm~\ref{alg:fpReduction}):

\begin{equation}
  \label{eq:CalculateScore1}
  AS_{ind-PAG} = N_s^{2} \times N_e^{2}
\end{equation}

where $N_s$ and $N_e$ are the degree of indirection in persistence setup atomic graph and persistence execution atomic graph, respectively. 

Second, the anomaly score of an indicator observable from the dependency graph is calculated as follows (Line~\ref{line:CalculateScore2} in Algorithm~\ref{alg:fpReduction}):

\begin{equation}
  \label{eq:CalculateScore2}
  \resizebox{.91\hsize}{!}{$
  AS_{ind-DG} = 
  \begin{cases}
  \max\limits_i (\frac{D_{c}}{{D_s}} \times Freq({teq_i}) \times Var({tac})) & t_{teq} \leq t_{e} \\
  \max\limits_i (\frac{D_{c}}{{D_e}} \times Freq({teq_i}) \times Var({tac})) & t_{teq} > t_{e} 
  \end{cases}$}
\end{equation}

where $D_s$ denotes the distance between an indicative attack step, e.g., OS credentials dumping, and the persistence setup step. The distance is measured as the number of hops between the corresponding two processes. Similarly, $D_e$ denotes the distance between the persistence execution step and an indicative attack step, e.g., remote system discovery. $D_{c}$ is a predefined cut-off value representing the maximal number of hops considered as having positive impact on the anomaly score. By doing so, it penalizes an indicative attack step too far away from the persistence setup or execution step, in which the weighting factor in Equation~\ref{eq:CalculateScore2} is less than 1, when $D_s$ or $D_e$ is greater than $D_{c}$. $Freq({teq})$ represents the occurrence of the same attack technique being executed as repeated attempt, whereas $Var({tac})$ represents usage extensiveness of the same tactic, i.e., number of different techniques from the same tactic being applied. The rationale behind this is that attackers often try a variety of techniques from the same tactic together to maximize the chance of achieving their objectives. $t_{teq}$ is the time when an attack step is conducted, and $t_{e}$ is the time when persistence execution is performed. If multiple attack \mbox{(sub-)techniques} are observed for the same indicator, we only consider the one with the maximal anomaly score. 

In the end, we obtain the final anomaly score from the Equation~\ref{eq:SumScore} (Line~\ref{line:SumScore} in Algorithm~\ref{alg:fpReduction}).

\begin{equation}
  \label{eq:SumScore}
  AS_{PE} = \prod_{i=1}^{n} (AS_i)^{w_i}
\end{equation}

where $n$ denotes the number of found indicators for a given \Edge, $AS_i$ denotes the anomaly score of an indicator obtained from Equation~\ref{eq:CalculateScore1} or Equation~\ref{eq:CalculateScore2} for this \Edge, and $w_i$ is a weighting factor. Afterwards, the \Edge and anomaly score pair is added to a list (Line~\ref{line:AddToList} in Algorithm~\ref{alg:fpReduction}), which is sorted by anomaly score at the end (Line~\ref{line:SortByScore} in Algorithm~\ref{alg:fpReduction}). \EDges ranked lower than the \bigNI{}-th \Edge are considered as false-positive \Edges, and therefore removed from the list (Line~\ref{line:RemoveByBudget} in Algorithm~\ref{alg:fpReduction}). Algorithm \ref{alg:fpReduction} returns the final list of \Edge and anomaly score pairs.

\section{Implementation}
For our experiments, we implemented the prototype of \Sys in Python ($\sim$ 6K lines of code), and deployed it on a 64bit Ubuntu 23.04 OS with 512 GB of RAM and a 64-core AMD processor. This machine hosts a dozen virtual machines used by several researchers for conducting separated scientific experiments at the same time. Our implementation interfaces Elasticsearch~\cite{elasticsearch} via EQL, a query language specifically designed for security use cases. Elasticsearch provides scalable and near real-time search for log data investigation. Besides, we use Python NetworkX~\cite{networkx} for generating provenance graphs on demand, and PyVis~\cite{pyvis} for graph visualization.

For Linux log collection, we use Auditd~\cite{linuxaudit}. On Windows, our primary tool is System Monitor (Sysmon)~\cite{sysmon}, which, unlike Windows ETW, generates and records a process GUID for each process, reducing false dependencies during post-processing. However, Sysmon lacks file/Registry read collection, so we use Windows Security Audit logs for recording all file and Registry operation. We also collect ALPC logs with the \quotes{NT Kernel Logger}~\cite{NTKernelLogger} ETW session.

\section{Evaluation}
\label{s:eval}
In this section, we evaluate the efficacy and effectiveness of \Sys as a persistence detection system. In particular, we investigate the following research questions (RQs):

\begin{itemize}[topsep=.2ex,itemsep=.2ex,leftmargin=0.9cm]
\item[\textbf{\emph{RQ{1}}}] How does \Sys compare in soundness of persistence detection, false positive reduction, and accuracy to open-source SIEM detection rules, commercial EDR systems and state-of-the-art PIDS? (\ref{ss:OpenSourceDetectionRules})
\item[\textbf{\emph{RQ{2}}}] How high is the log reduction rate by introducing expert-guided edges in \Sys? (\ref{ss:logReduction})
\item[\textbf{\emph{RQ{3}}}] What is the runtime overhead of \Sys? (\ref{ss:ResponseTime})
\item[\textbf{\emph{RQ{4}}}] How precise are persistence attack graphs generated by \Sys? (\ref{ss:AttackGraph})
\end{itemize}

\PP{Public Datasets} Public datasets often lack persistence traces. From DARPA datasets, we chose the E5 dataset~\cite{TransparentComputing} and OpTC dataset~\cite{OpTC}, but omitted the E3~\cite{TransparentComputingE3} due to its lack of persistence attacks. DARPA E5 dataset features emulated APT attacks. Only the E5 Fivedirections subset, focused on Windows, contains two instances of persistence attacks. We evaluated only this subset with \Sys. The DARPA OpTC dataset contains logs from 500 Windows machines. It includes three persistence instances, but only one meets all criteria from Section~\ref{ss:motivation}. The other two failed the third condition as the attack ended prematurely. We tested \Sys on a subset of 50 machines, including the 3 with persistence and 47 random ones. We also included ATLASv2~\cite{ATLASv2} for its CBC detection results on persistence. ATLASv2 dataset offers more background activities and extensive logging than ATLAS~\cite{atlas}, such as through Sysmon and VMware CBC~\cite{carbonblackcloud}. While lacking actual persistence attacks, it includes CBC's persistence alerts. We used these for comparison with \Sys. \Sys's detection results on these datasets are verified against the provided ground truth.

\PP{MITRE Attack Emulation} MITRE's eleven full emulation plans~\cite{AdversaryEmulationLibrary}, based on real APT behaviors, each include at least two persistence techniques. These plans are used in MITRE Engenuity ATT\&CK\textsuperscript{®} Evaluations~\cite{MITREEngenuity} to assess commercial EDR systems~\cite{MITREEngenuityEvaluation}. However, MITRE has not published any corresponding datasets. We precisely implemented two relevant emulation plans, focusing on top 10 most \quotes{persistent} APT groups from Section \ref{ss:PersistencePrevalence}, and then evaluated \Sys on these datasets, valuable for PIDS research. MITRE's emulation plans target enterprise networks with more sophisticated, cross-machine attacks than most public datasets, offering greater authenticity. The APT29 plan has two distinct scenarios, while Sandworm's are identical. We created three datasets from emulating APT29 scenario 1, APT29 scenario 2, and Sandworm scenario 1, named EP-APT29-1, EP-APT29-2, and EP-Sandworm-1, respectively. Table~\ref{tab:evaluation:datasetsOverview} gives an overview of our datasets.

\begin{table}[!tbp]
\centering
  \caption{Overview of the evaluation datasets}
  \resizebox{0.48\textwidth}{!}{
  \begin{tabular}{cccccccc}
    \toprule
    \makecell{Dataset} & \makecell{Target Host\\Number} & \makecell{Persistence\\Attack Number}  & \makecell{Target\\Host OS} & \makecell{Data\\Size}  & \makecell{Event\\Number}  \\
    \midrule
     ATLASv2 & 2 & 0 & Windows & 26GB  & 5.6M  \\
    \midrule
    \makecell{DARPA-E5-\\Fivediretions} & 3 & 2 & Windows & 348GB &  1.4B  \\
    \midrule
    \makecell{DARPA-\\OpTC} & \makecell{50\\(/500)} & 1 & Windows & 380GB & 338M   \\
    \midrule
     EP-APT29-1 & 3 & 2 & Windows & 32GB &  22M \\
    \midrule
     EP-APT29-2 & 3 & 3 & Windows & 24GB &  14M   \\
    \midrule
     \makecell{EP-\\Sandworm-1} & 4 & 4 & \makecell{Windows\\Linux}  & 68GB & 57M  \\
    \bottomrule
  \end{tabular}}
  \label{tab:evaluation:datasetsOverview}
\end{table}

\subsection{Effectiveness of \Sys}
\label{ss:OpenSourceDetectionRules}
{\renewcommand{\arraystretch}{1.2}
\begin{table*}[!t]
  \centering
  \scriptsize
  \caption{Comparison of \Sys and open-source SIEM rules.}
  \begin{threeparttable}
  \begin{tabular}{ccccccccccccc}
    \toprule

  \multirow{3}{*}{\textbf{Datasets}}
  & \multicolumn{6}{c }{\Norothead{ \bf \Sys}}
  & \multicolumn{2}{c }{\Norothead{ \bf Elastic }}
  & \multicolumn{2}{c }{\Norothead{ \bf Chronicle }}
  & \multicolumn{2}{c }{\Norothead{ \bf Sigma }}

  \\ \cmidrule(r{\tbspace}){2-7} 
  
  \multirow{1}{*}{\textbf{}} & \multicolumn{2}{c }{\Norothead{ \bf Stage 1}} & \multicolumn{2}{c }{\Norothead{ \bf Stage 2 }} & \multicolumn{2}{c }{\Norothead{ \bf Stage 3 }}
    
  \\ \cmidrule(r{\tbspace}){2-3} \cmidrule(r{\tbspace}){4-5} \cmidrule(r{\tbspace}){6-7} \cmidrule(r{\tbspace}){8-9} \cmidrule(r{\tbspace}){10-11} \cmidrule(r{\tbspace}){12-13}  

    &  {\bf FP} & {\bf FN}  & {\bf FP} & {\bf FN} & {\bf FP}  & {\bf FN} & {\bf FP} & {\bf FN} & {\bf FP} & {\bf FN} & {\bf FP}  & {\bf FN} \\

  \midrule

  DARPA-E5 & \FPeval{\result}{clip(\UseMacro{DARPA_E5_FD1_COMMANDER_Stage1_FP}+\UseMacro{DARPA_E5_FD2_COMMANDER_Stage1_FP}+\UseMacro{DARPA_E5_FD3_COMMANDER_Stage1_FP})}$\result$ & \FPeval{\result}{clip(\UseMacro{DARPA_E5_FD1_COMMANDER_Stage1_FN}+\UseMacro{DARPA_E5_FD2_COMMANDER_Stage1_FN}+\UseMacro{DARPA_E5_FD3_COMMANDER_Stage1_FN})}$\result$/2\tnote{*} & \FPeval{\result}{clip(\UseMacro{DARPA_E5_FD1_COMMANDER_Stage2_FP}+\UseMacro{DARPA_E5_FD2_COMMANDER_Stage2_FP}+\UseMacro{DARPA_E5_FD3_COMMANDER_Stage2_FP})}$\result$ & \FPeval{\result}{clip(\UseMacro{DARPA_E5_FD1_COMMANDER_Stage2_FN}+\UseMacro{DARPA_E5_FD2_COMMANDER_Stage2_FN}+\UseMacro{DARPA_E5_FD3_COMMANDER_Stage2_FN})}$\result$/2 & \UseMacro{DARPA_E5_COMMANDER_Stage3_FP}  & \UseMacro{DARPA_E5_COMMANDER_Stage3_FN}/2 & \FPeval{\result}{clip(\UseMacro{DARPA_E5_FD1_Elastic_FP}+\UseMacro{DARPA_E5_FD2_Elastic_FP}+\UseMacro{DARPA_E5_FD3_Elastic_FP})}$\result$ & \FPeval{\result}{clip(\UseMacro{DARPA_E5_FD1_Elastic_FN}+\UseMacro{DARPA_E5_FD2_Elastic_FN}+\UseMacro{DARPA_E5_FD3_Elastic_FN})}$\result$/2 & - & - & \FPeval{\result}{clip(\UseMacro{DARPA_E5_FD1_Sigma_FP}+\UseMacro{DARPA_E5_FD2_Sigma_FP}+\UseMacro{DARPA_E5_FD3_Sigma_FP})}$\result$ & \FPeval{\result}{clip(\UseMacro{DARPA_E5_FD1_Sigma_FN}+\UseMacro{DARPA_E5_FD2_Sigma_FN}+\UseMacro{DARPA_E5_FD3_Sigma_FN})}$\result$/2\\
  
  \midrule

  DARPA-OpTC & \UseMacro{DARPA_OpTC_COMMANDER_Stage1_FP} & \UseMacro{DARPA_OpTC_COMMANDER_Stage1_FN}/1 & \UseMacro{DARPA_OpTC_COMMANDER_Stage2_FP} & \UseMacro{DARPA_OpTC_COMMANDER_Stage2_FN}/1 & \UseMacro{DARPA_OpTC_COMMANDER_Stage3_FP}  & \UseMacro{DARPA_OpTC_COMMANDER_Stage3_FN}/1 &  \UseMacro{DARPA_OpTC_Elastic_FP} & \UseMacro{DARPA_OpTC_Elastic_FN}/1 & \UseMacro{DARPA_OpTC_Chronicle_FP} & \UseMacro{DARPA_OpTC_Chronicle_FN}/1 & \UseMacro{DARPA_OpTC_Sigma_FP} & \UseMacro{DARPA_OpTC_Sigma_FN}/1\\

 \midrule

  EP-APT29-1 & \FPeval{\result}{clip(\UseMacro{APT29_Scenario1_g3mef76_malicious_COMMANDER_Stage1_FP}+\UseMacro{APT29_Scenario1_g3mef77_malicious_COMMANDER_Stage1_FN}+\UseMacro{APT29_Scenario1_rc1r6g07ik7_malicious_COMMANDER_Stage1_FP}+\UseMacro{APT29_Scenario1_g3mef76_benign_COMMANDER_Stage1_FP}+\UseMacro{APT29_Scenario1_g3mef77_benign_COMMANDER_Stage1_FN}+\UseMacro{APT29_Scenario1_rc1r6g07ik7_benign_COMMANDER_Stage1_FP})}$\result$  & \FPeval{\result}{clip(\UseMacro{APT29_Scenario1_g3mef76_malicious_COMMANDER_Stage1_FN}+\UseMacro{APT29_Scenario1_g3mef77_malicious_COMMANDER_Stage1_FN}+\UseMacro{APT29_Scenario1_rc1r6g07ik7_malicious_COMMANDER_Stage1_FN}+\UseMacro{APT29_Scenario1_g3mef76_benign_COMMANDER_Stage1_FN}+\UseMacro{APT29_Scenario1_g3mef77_benign_COMMANDER_Stage1_FN}+\UseMacro{APT29_Scenario1_rc1r6g07ik7_benign_COMMANDER_Stage1_FN})}$\result$/2 & \FPeval{\result}{clip(\UseMacro{APT29_Scenario1_g3mef76_malicious_COMMANDER_Stage2_FP}+\UseMacro{APT29_Scenario1_g3mef77_malicious_COMMANDER_Stage2_FP}+\UseMacro{APT29_Scenario1_rc1r6g07ik7_malicious_COMMANDER_Stage2_FP}+\UseMacro{APT29_Scenario1_g3mef76_benign_COMMANDER_Stage2_FP}+\UseMacro{APT29_Scenario1_g3mef77_benign_COMMANDER_Stage2_FP}+\UseMacro{APT29_Scenario1_rc1r6g07ik7_benign_COMMANDER_Stage2_FP})}$\result$ & \FPeval{\result}{clip(\UseMacro{APT29_Scenario1_g3mef76_malicious_COMMANDER_Stage2_FN}+\UseMacro{APT29_Scenario1_g3mef77_malicious_COMMANDER_Stage2_FN}+\UseMacro{APT29_Scenario1_rc1r6g07ik7_malicious_COMMANDER_Stage2_FN}+\UseMacro{APT29_Scenario1_g3mef76_benign_COMMANDER_Stage2_FN}+\UseMacro{APT29_Scenario1_g3mef77_benign_COMMANDER_Stage2_FN}+\UseMacro{APT29_Scenario1_rc1r6g07ik7_benign_COMMANDER_Stage2_FN})}$\result$/2 & \UseMacro{APT29_Scenario1_COMMANDER_Stage3_FP}  & \UseMacro{APT29_Scenario1_COMMANDER_Stage3_FN}/2 &  \FPeval{\result}{clip(\UseMacro{APT29_Scenario1_g3mef76_malicious_Elastic_FP}+\UseMacro{APT29_Scenario1_g3mef77_malicious_Elastic_FP}+\UseMacro{APT29_Scenario1_rc1r6g07ik7_malicious_Elastic_FP}+\UseMacro{APT29_Scenario1_g3mef76_benign_Elastic_FP}+\UseMacro{APT29_Scenario1_g3mef77_benign_Elastic_FP}+\UseMacro{APT29_Scenario1_rc1r6g07ik7_benign_Elastic_FP})}$\result$ & \FPeval{\result}{clip(\UseMacro{APT29_Scenario1_g3mef76_malicious_Elastic_FN}+\UseMacro{APT29_Scenario1_g3mef77_malicious_Elastic_FN}+\UseMacro{APT29_Scenario1_rc1r6g07ik7_malicious_Elastic_FN}+\UseMacro{APT29_Scenario1_g3mef76_benign_Elastic_FN}+\UseMacro{APT29_Scenario1_g3mef77_benign_Elastic_FN}+\UseMacro{APT29_Scenario1_rc1r6g07ik7_benign_Elastic_FN})}$\result$/2 & \FPeval{\result}{clip(\UseMacro{APT29_Scenario1_g3mef76_malicious_Chronicle_FP}+\UseMacro{APT29_Scenario1_g3mef77_malicious_Chronicle_FP}+\UseMacro{APT29_Scenario1_rc1r6g07ik7_malicious_Chronicle_FP}+\UseMacro{APT29_Scenario1_g3mef76_benign_Chronicle_FP}+\UseMacro{APT29_Scenario1_g3mef77_benign_Chronicle_FP}+\UseMacro{APT29_Scenario1_rc1r6g07ik7_benign_Chronicle_FP})}$\result$ & \FPeval{\result}{clip(\UseMacro{APT29_Scenario1_g3mef76_malicious_Chronicle_FN}+\UseMacro{APT29_Scenario1_g3mef77_malicious_Chronicle_FN}+\UseMacro{APT29_Scenario1_rc1r6g07ik7_malicious_Chronicle_FN}+\UseMacro{APT29_Scenario1_g3mef76_benign_Chronicle_FN}+\UseMacro{APT29_Scenario1_g3mef77_benign_Chronicle_FN}+\UseMacro{APT29_Scenario1_rc1r6g07ik7_benign_Chronicle_FN})}$\result$/2 & \FPeval{\result}{clip(\UseMacro{APT29_Scenario1_g3mef76_malicious_Sigma_FP}+\UseMacro{APT29_Scenario1_g3mef77_malicious_Sigma_FP}+\UseMacro{APT29_Scenario1_rc1r6g07ik7_malicious_Sigma_FP}+\UseMacro{APT29_Scenario1_g3mef76_benign_Sigma_FP}+\UseMacro{APT29_Scenario1_g3mef77_benign_Sigma_FP}+\UseMacro{APT29_Scenario1_rc1r6g07ik7_benign_Sigma_FP})}$\result$ & \FPeval{\result}{clip(\UseMacro{APT29_Scenario1_g3mef76_malicious_Sigma_FN}+\UseMacro{APT29_Scenario1_g3mef77_malicious_Sigma_FN}+\UseMacro{APT29_Scenario1_rc1r6g07ik7_malicious_Sigma_FN}+\UseMacro{APT29_Scenario1_g3mef76_benign_Sigma_FN}+\UseMacro{APT29_Scenario1_g3mef77_benign_Sigma_FN}+\UseMacro{APT29_Scenario1_rc1r6g07ik7_benign_Sigma_FN})}$\result$/2  \\
 
  \midrule
  
  EP-APT29-2 & \FPeval{\result}{clip(\UseMacro{APT29_Scenario2_g3mef76_malicious_COMMANDER_Stage1_FP}+\UseMacro{APT29_Scenario2_g3mef77_malicious_COMMANDER_Stage1_FN}+\UseMacro{APT29_Scenario2_rc1r6g07ik7_malicious_COMMANDER_Stage1_FP}+\UseMacro{APT29_Scenario2_g3mef76_benign_COMMANDER_Stage1_FP}+\UseMacro{APT29_Scenario2_g3mef77_benign_COMMANDER_Stage1_FN}+\UseMacro{APT29_Scenario2_rc1r6g07ik7_benign_COMMANDER_Stage1_FP})}$\result$  & \FPeval{\result}{clip(\UseMacro{APT29_Scenario2_g3mef76_malicious_COMMANDER_Stage1_FN}+\UseMacro{APT29_Scenario2_g3mef77_malicious_COMMANDER_Stage1_FN}+\UseMacro{APT29_Scenario2_rc1r6g07ik7_malicious_COMMANDER_Stage1_FN}+\UseMacro{APT29_Scenario2_g3mef76_benign_COMMANDER_Stage1_FN}+\UseMacro{APT29_Scenario2_g3mef77_benign_COMMANDER_Stage1_FN}+\UseMacro{APT29_Scenario2_rc1r6g07ik7_benign_COMMANDER_Stage1_FN})}$\result$/3 & \FPeval{\result}{clip(\UseMacro{APT29_Scenario2_g3mef76_malicious_COMMANDER_Stage2_FP}+\UseMacro{APT29_Scenario2_g3mef77_malicious_COMMANDER_Stage2_FP}+\UseMacro{APT29_Scenario2_rc1r6g07ik7_malicious_COMMANDER_Stage2_FP}+\UseMacro{APT29_Scenario2_g3mef76_benign_COMMANDER_Stage2_FP}+\UseMacro{APT29_Scenario2_g3mef77_benign_COMMANDER_Stage2_FP}+\UseMacro{APT29_Scenario2_rc1r6g07ik7_benign_COMMANDER_Stage2_FP})}$\result$ & \FPeval{\result}{clip(\UseMacro{APT29_Scenario2_g3mef76_malicious_COMMANDER_Stage2_FN}+\UseMacro{APT29_Scenario2_g3mef77_malicious_COMMANDER_Stage2_FN}+\UseMacro{APT29_Scenario2_rc1r6g07ik7_malicious_COMMANDER_Stage2_FN}+\UseMacro{APT29_Scenario2_g3mef76_benign_COMMANDER_Stage2_FN}+\UseMacro{APT29_Scenario2_g3mef77_benign_COMMANDER_Stage2_FN}+\UseMacro{APT29_Scenario2_rc1r6g07ik7_benign_COMMANDER_Stage2_FN})}$\result$/3 & \UseMacro{APT29_Scenario2_COMMANDER_Stage3_FP}  & \UseMacro{APT29_Scenario2_COMMANDER_Stage3_FN}/3 &  \FPeval{\result}{clip(\UseMacro{APT29_Scenario2_g3mef76_malicious_Elastic_FP}+\UseMacro{APT29_Scenario2_g3mef77_malicious_Elastic_FP}+\UseMacro{APT29_Scenario2_rc1r6g07ik7_malicious_Elastic_FP}+\UseMacro{APT29_Scenario2_g3mef76_benign_Elastic_FP}+\UseMacro{APT29_Scenario2_g3mef77_benign_Elastic_FP}+\UseMacro{APT29_Scenario2_rc1r6g07ik7_benign_Elastic_FP})}$\result$ & \FPeval{\result}{clip(\UseMacro{APT29_Scenario2_g3mef76_malicious_Elastic_FN}+\UseMacro{APT29_Scenario2_g3mef77_malicious_Elastic_FN}+\UseMacro{APT29_Scenario2_rc1r6g07ik7_malicious_Elastic_FN}+\UseMacro{APT29_Scenario2_g3mef76_benign_Elastic_FN}+\UseMacro{APT29_Scenario2_g3mef77_benign_Elastic_FN}+\UseMacro{APT29_Scenario2_rc1r6g07ik7_benign_Elastic_FN})}$\result$/3 & \FPeval{\result}{clip(\UseMacro{APT29_Scenario2_g3mef76_malicious_Chronicle_FP}+\UseMacro{APT29_Scenario2_g3mef77_malicious_Chronicle_FP}+\UseMacro{APT29_Scenario2_rc1r6g07ik7_malicious_Chronicle_FP}+\UseMacro{APT29_Scenario2_g3mef76_benign_Chronicle_FP}+\UseMacro{APT29_Scenario2_g3mef77_benign_Chronicle_FP}+\UseMacro{APT29_Scenario2_rc1r6g07ik7_benign_Chronicle_FP})}$\result$ & \FPeval{\result}{clip(\UseMacro{APT29_Scenario2_g3mef76_malicious_Chronicle_FN}+\UseMacro{APT29_Scenario2_g3mef77_malicious_Chronicle_FN}+\UseMacro{APT29_Scenario2_rc1r6g07ik7_malicious_Chronicle_FN}+\UseMacro{APT29_Scenario2_g3mef76_benign_Chronicle_FN}+\UseMacro{APT29_Scenario2_g3mef77_benign_Chronicle_FN}+\UseMacro{APT29_Scenario2_rc1r6g07ik7_benign_Chronicle_FN})}$\result$/3  &\FPeval{\result}{clip(\UseMacro{APT29_Scenario2_g3mef76_malicious_Sigma_FP}+\UseMacro{APT29_Scenario2_g3mef77_malicious_Sigma_FP}+\UseMacro{APT29_Scenario2_rc1r6g07ik7_malicious_Sigma_FP}+\UseMacro{APT29_Scenario2_g3mef76_benign_Sigma_FP}+\UseMacro{APT29_Scenario2_g3mef77_benign_Sigma_FP}+\UseMacro{APT29_Scenario2_rc1r6g07ik7_benign_Sigma_FP})}$\result$ & \FPeval{\result}{clip(\UseMacro{APT29_Scenario2_g3mef76_malicious_Sigma_FN}+\UseMacro{APT29_Scenario2_g3mef77_malicious_Sigma_FN}+\UseMacro{APT29_Scenario2_rc1r6g07ik7_malicious_Sigma_FN}+\UseMacro{APT29_Scenario2_g3mef76_benign_Sigma_FN}+\UseMacro{APT29_Scenario2_g3mef77_benign_Sigma_FN}+\UseMacro{APT29_Scenario2_rc1r6g07ik7_benign_Sigma_FN})}$\result$/3 \\
  
  \midrule

  EP-Sandworm-1 & \FPeval{\result}{clip(\UseMacro{Sandworm_Scenario1_g3mef76_malicious_COMMANDER_Stage1_FP}+\UseMacro{Sandworm_Scenario1_g3mef77_malicious_COMMANDER_Stage1_FN}+\UseMacro{Sandworm_Scenario1_rc1r6g07ik7_malicious_COMMANDER_Stage1_FP}+\UseMacro{Sandworm_Scenario1_g3mef76_benign_COMMANDER_Stage1_FP}+\UseMacro{Sandworm_Scenario1_g3mef77_benign_COMMANDER_Stage1_FN}+\UseMacro{Sandworm_Scenario1_rc1r6g07ik7_benign_COMMANDER_Stage1_FP}+\UseMacro{Sandworm_Scenario1_malicious_Linux_COMMANDER_Stage1_FP}+\UseMacro{Sandworm_Scenario1_benign_Linux_COMMANDER_Stage1_FP})}$\result$ & \FPeval{\result}{clip(\UseMacro{Sandworm_Scenario1_g3mef76_malicious_COMMANDER_Stage1_FN}+\UseMacro{Sandworm_Scenario1_g3mef77_malicious_COMMANDER_Stage1_FN}+\UseMacro{Sandworm_Scenario1_rc1r6g07ik7_malicious_COMMANDER_Stage1_FN}+\UseMacro{Sandworm_Scenario1_g3mef76_benign_COMMANDER_Stage1_FN}+\UseMacro{Sandworm_Scenario1_g3mef77_benign_COMMANDER_Stage1_FN}+\UseMacro{Sandworm_Scenario1_rc1r6g07ik7_benign_COMMANDER_Stage1_FN}+\UseMacro{Sandworm_Scenario1_malicious_Linux_COMMANDER_Stage1_FN}+\UseMacro{Sandworm_Scenario1_benign_Linux_COMMANDER_Stage1_FN})}$\result$/4 & \FPeval{\result}{clip(\UseMacro{Sandworm_Scenario1_g3mef76_malicious_COMMANDER_Stage2_FP}+\UseMacro{Sandworm_Scenario1_g3mef77_malicious_COMMANDER_Stage2_FP}+\UseMacro{Sandworm_Scenario1_rc1r6g07ik7_malicious_COMMANDER_Stage2_FP}+\UseMacro{Sandworm_Scenario1_g3mef76_benign_COMMANDER_Stage2_FP}+\UseMacro{Sandworm_Scenario1_g3mef77_benign_COMMANDER_Stage2_FP}+\UseMacro{Sandworm_Scenario1_rc1r6g07ik7_benign_COMMANDER_Stage2_FP}+\UseMacro{Sandworm_Scenario1_malicious_Linux_COMMANDER_Stage2_FP}+\UseMacro{Sandworm_Scenario1_benign_Linux_COMMANDER_Stage2_FP})}$\result$ & \FPeval{\result}{clip(\UseMacro{Sandworm_Scenario1_g3mef76_malicious_COMMANDER_Stage2_FN}+\UseMacro{Sandworm_Scenario1_g3mef77_malicious_COMMANDER_Stage2_FN}+\UseMacro{Sandworm_Scenario1_rc1r6g07ik7_malicious_COMMANDER_Stage2_FN}+\UseMacro{Sandworm_Scenario1_g3mef76_benign_COMMANDER_Stage2_FN}+\UseMacro{Sandworm_Scenario1_g3mef77_benign_COMMANDER_Stage2_FN}+\UseMacro{Sandworm_Scenario1_rc1r6g07ik7_benign_COMMANDER_Stage2_FN}+\UseMacro{Sandworm_Scenario1_malicious_Linux_COMMANDER_Stage2_FN}+\UseMacro{Sandworm_Scenario1_benign_Linux_COMMANDER_Stage2_FN})}$\result$/4 & \FPeval{\result}{clip(\UseMacro{Sandworm_Scenario1_Win_COMMANDER_Stage3_FP}+\UseMacro{Sandworm_Scenario1_Linux_COMMANDER_Stage3_FP})}$\result$  & \FPeval{\result}{clip(\UseMacro{Sandworm_Scenario1_Win_COMMANDER_Stage3_FN}+\UseMacro{Sandworm_Scenario1_Linux_COMMANDER_Stage3_FN})}$\result$/4  &  \FPeval{\result}{clip(\UseMacro{Sandworm_Scenario1_g3mef76_malicious_Elastic_FP}+\UseMacro{Sandworm_Scenario1_g3mef77_malicious_Elastic_FP}+\UseMacro{Sandworm_Scenario1_rc1r6g07ik7_malicious_Elastic_FP}+\UseMacro{Sandworm_Scenario1_g3mef76_benign_Elastic_FP}+\UseMacro{Sandworm_Scenario1_g3mef77_benign_Elastic_FP}+\UseMacro{Sandworm_Scenario1_rc1r6g07ik7_benign_Elastic_FP})}$\result$ & \FPeval{\result}{clip(\UseMacro{Sandworm_Scenario1_g3mef76_malicious_Elastic_FN}+\UseMacro{Sandworm_Scenario1_g3mef77_malicious_Elastic_FN}+\UseMacro{Sandworm_Scenario1_rc1r6g07ik7_malicious_Elastic_FN}+\UseMacro{Sandworm_Scenario1_g3mef76_benign_Elastic_FN}+\UseMacro{Sandworm_Scenario1_g3mef77_benign_Elastic_FN}+\UseMacro{Sandworm_Scenario1_rc1r6g07ik7_benign_Elastic_FN})}$\result$/4 & \FPeval{\result}{clip(\UseMacro{Sandworm_Scenario1_g3mef76_malicious_Chronicle_FP}+\UseMacro{Sandworm_Scenario1_g3mef77_malicious_Chronicle_FP}+\UseMacro{Sandworm_Scenario1_rc1r6g07ik7_malicious_Chronicle_FP}+\UseMacro{Sandworm_Scenario1_g3mef76_benign_Chronicle_FP}+\UseMacro{Sandworm_Scenario1_g3mef77_benign_Chronicle_FP}+\UseMacro{Sandworm_Scenario1_rc1r6g07ik7_benign_Chronicle_FP})}$\result$ & \FPeval{\result}{clip(\UseMacro{Sandworm_Scenario1_g3mef76_malicious_Chronicle_FN}+\UseMacro{Sandworm_Scenario1_g3mef77_malicious_Chronicle_FN}+\UseMacro{Sandworm_Scenario1_rc1r6g07ik7_malicious_Chronicle_FN}+\UseMacro{Sandworm_Scenario1_g3mef76_benign_Chronicle_FN}+\UseMacro{Sandworm_Scenario1_g3mef77_benign_Chronicle_FN}+\UseMacro{Sandworm_Scenario1_rc1r6g07ik7_benign_Chronicle_FN})}$\result$/4 & \FPeval{\result}{clip(\UseMacro{Sandworm_Scenario1_g3mef76_malicious_Sigma_FP}+\UseMacro{Sandworm_Scenario1_g3mef77_malicious_Sigma_FP}+\UseMacro{Sandworm_Scenario1_rc1r6g07ik7_malicious_Sigma_FP}+\UseMacro{Sandworm_Scenario1_g3mef76_benign_Sigma_FP}+\UseMacro{Sandworm_Scenario1_g3mef77_benign_Sigma_FP}+\UseMacro{Sandworm_Scenario1_rc1r6g07ik7_benign_Sigma_FP})}$\result$ & \FPeval{\result}{clip(\UseMacro{Sandworm_Scenario1_g3mef76_malicious_Sigma_FN}+\UseMacro{Sandworm_Scenario1_g3mef77_malicious_Sigma_FN}+\UseMacro{Sandworm_Scenario1_rc1r6g07ik7_malicious_Sigma_FN}+\UseMacro{Sandworm_Scenario1_g3mef76_benign_Sigma_FN}+\UseMacro{Sandworm_Scenario1_g3mef77_benign_Sigma_FN}+\UseMacro{Sandworm_Scenario1_rc1r6g07ik7_benign_Sigma_FN})}$\result$/4 \\

  \bottomrule
  \end{tabular}
  \begin{tablenotes}\footnotesize
   \item[*] The number before / represents false negatives, and the number after / represents true positives in the corresponding dataset.
   \end{tablenotes}
   \end{threeparttable}
\label{tab:evaluation:results}
\end{table*}

}

\begin{figure*}[!t]
  \centering
  \includegraphics[width=0.85\linewidth]{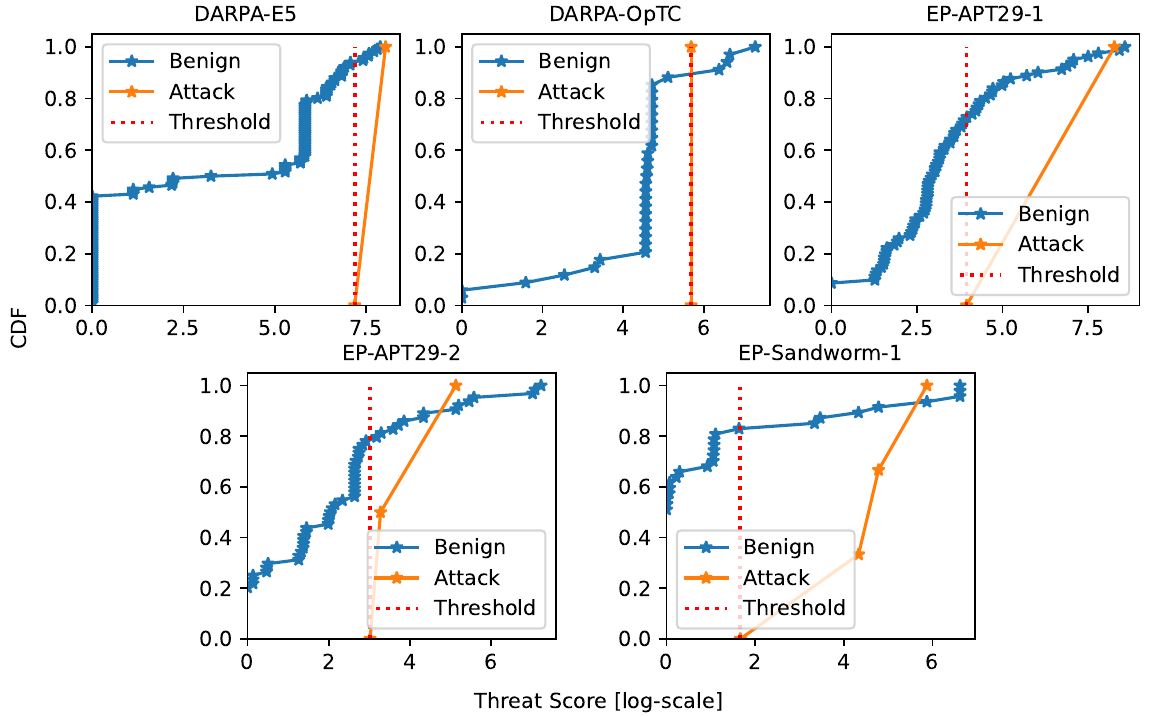}
  \caption{CDF of threat score for false and true alerts}
  \label{fig:ThreatScore}
  \vspace{-2ex}
\end{figure*}

\PP{\Sys vs. SIEM detection rules}
After finalizing our persistence detection rule set, we compared it with popular open-source SIEM detection rules from Elastic~\cite{elasticdetectionrules}, Sigma~\cite{sigma} and Google Chronicle~\cite{chronicledetectionrules}. We extracted all persistence-related
detection rules from these repositories, converted them into EQL queries and ran them alongside \Sys on datasets containing persistence attacks. The results in Table~\ref{tab:evaluation:results} reveal that open-source SIEM rules generated more false positives and missed true attacks, and \Sys significantly outperformed these rules. Further comparison between \Sys's stages 2 and 3 is presented in Figure~\ref{fig:ThreatScore}. This figure displays the cumulative distribution of threat scores for benign and attack \Edges. The stage 3 results in Table~\ref{tab:evaluation:results} are based on the lowest true attack threat score threshold. \Sys's final stage drastically reduced false positives—from thousands in open-source rules to just dozens per dataset. Additionally, the false positive reduction algorithm in stage 3 of \Sys achieved an average reduction rate of \UseMacro{fpr_avg_2to3_reduction}\%, significantly enhancing accuracy.

\Sys's great improvement over SIEM detection rules mainly result from our two key insights, implemented on its stage 2 and 3 respectively. Our first powerful insight is that all persistence attacks require a setup stage that (mis)uses a \quotes{sensitive} system functionality and a subsequent execution stage linked with a remote connection. Both SIEM rules and \Sys's stage 1 rules raise alerts when persistence-related system functionality is used, but not necessarily misused. Unlike SIEM rules, which don’t associate log events generated at different times, \Sys's stage 2 checks if there is a remote connection that can be traced back to a setup alert, leading to removal of alerts on system activities associated with benign usage of system functionality. Our second practical insight is that persistence is only one of multiple stages in a cyber-kill-chain that must be executed together to achieve attackers' goals. This key insight is discussed in details in our false positive reduction algorithm in Section~\ref{ss:FalsePositiveReduction}. Both key insights fully leverage the power of provenance analytics, i.e., context provided by system activities across a large timespan.

All parameters in our false positive reduction algorithms, e.g., weighting factors, are unchanged for each dataset. Like previous works ~\cite{holmes2019, nodoze2019, rapsheet2020}, our goal is not to exclude all FPs, but rather to prioritize most potential attacks for investigation using threat score ranking. In other words, \Sys aims to maximize the likelihood of detecting persistence attacks based on limited human resources. Based on the alert budget security analysts have, they can flexibly configure this parameter. 

We further investigate on why these open-source detection rules have undesired outcomes, in particular, 1) why there are many false negatives when applying Elastic's rules; 2) why Sigma's rules create exceptionally high number of false positives. To answer the first question, we carefully inspect Elastic's persistence detection rules. We find out that Elastic's rules not only overly allowlist programs, but also tend to be too specific, containing many hard-coded strings as conditions. Both cases make them very vulnerable to evasion attacks and have many false negatives in practice. For instance, this Elastic rule for T1547.001~\cite{ElasticSample} allowlists all programs not only under \texttt{C:\textbackslash Windows\textbackslash System32\textbackslash} but also under \texttt{C:\textbackslash Program Files\textbackslash}, essentially excluding the majority of programs on Windows. As discussed above, DARPA OpTC has three persistence instances, including two \quotes{unsuccessful} persistence instances only due to incomplete attack scenarios.  We find that, due to the high-degree of specificity and overly allowlisting, Elastic's rules missed all these three persistence instances. On the contrary, our system \Sys detected all three persistence instances in stage 1, and removed the two unsuccessful persistence instances in stage 2, presenting only the true persistence as its final output.

To answer the second question, we resort to manually inspecting Sigma's persistence detection rules. We find that Sigma's rules are more diverse, with some being overly specific and others being too general. This Sigma rule sample for T1574.009~\cite{SigmaSample} uses only a program file path as condition, leading to excessive amount of alerts. We also find that the high number of alerts is partly due to the fact that the Sigma rule repository contains many similar rules created by different contributors for the same attack \mbox{(sub-)techniques}. We argue that the repository maintainers should more properly organize the rules and remove seemingly redundant rules. 

In comparison to Sigma, Chronicle's rules cause less alerts, but have a few false negatives. Besides, we also find out that Chronicle has overly simple rules causing too many alerts such as this rule for T1053.005~\cite{ChronicleSample}. At last, Chronicle's rules overly use process names/paths as conditions. However, in the DARPA E5 dataset, process name is absent in log events caused by processes that were created before the logging framework started. Lacking this information will result in wrong results. Hence we do not evaluate Chronicle's rules on the DARPA E5 dataset.  It is worth mentioning that this study focuses on detecting persistence, which is often overlooked or even badly understood, as explained in Section~~\ref{ss:motivation}. It is not surprising that those popular detection rule repositories have less sound or complete \text{persistence} detection rule sets.

\PP{\Sys vs. CBC EDR}
We then compare our system \Sys with the commercial VMware CBC EDR~\cite{carbonblackcloud} on ATLASv2 dataset. For fair comparison, we did not run \Sys's stage 3 on this dataset, because it does not include a real persistence attack. Table~\ref{tab:evaluation:cbc_results} shows the detection results of \Sys and CBC EDR. Like above, without running its stage 3, \Sys already outperforms the CBC EDR by reducing the false positive rate by \UseMacro{fp_reduction_CBC}\%. Besides, we find that CBC's IOC (indicator of compromise) hits reveal that they blindly allowlist programs as well, like SIEM detection rules discussed above. This could easily result in false negatives.

We stress that \Sys's stage 1 also eliminates excessive false alarms caused by critical system programs modifying files and Registry at indicative locations. But \Sys first checks if those allowlisted programs are potentially compromised by examining whether their executable files are modified. To further optimize the detection results at the stage 1 of \Sys, it does not create unnecessary alerts for DLL files dropped on disk that get deleted afterwards. Otherwise it would produce lots of security alerts related to several persistence \mbox{(sub-)techniques}, e.g., T1574.001 and T1574.002. This is due to a common Windows program behavior, in which a program drops some DLL files to a (temporary) file folder after being started, then it starts some new instances (as child processes) that load those DLL files. The DLL files get deleted when those child processes terminate. However, unlike Linux, Windows by default never deletes temporary files, and leaves it to the programs for the clean-up. That is, temporary files persist reboots if not deleted by their creator. Hence, if a DLL file is dropped and not deleted afterwards, \Sys generates a persistence setup alert for it in its stage 1.

{\renewcommand{\arraystretch}{1.2}
\begin{table}[!t]
  \centering
  \scriptsize
  \caption{Comparison of \Sys and CBC EDR}
  \begin{tabular}{ccccccccc}
    \toprule

  \multirow{3}{*}{\textbf{Dataset}}
  & \multicolumn{6}{c }{\Norothead{ \bf \Sys}}
  & \multicolumn{2}{c }{\Norothead{ \bf CBC }}

  \\ \cmidrule(r{\tbspace}){2-7} 
  
  \multirow{1}{*}{\textbf{}} & \multicolumn{2}{c }{\Norothead{ \bf Stage 1}} & \multicolumn{2}{c }{\Norothead{ \bf Stage 2 }} & \multicolumn{2}{c }{\Norothead{ \bf Stage 3 }}
    
  \\ \cmidrule(r{\tbspace}){2-3} \cmidrule(r{\tbspace}){4-5} \cmidrule(r{\tbspace}){6-7} \cmidrule(r{\tbspace}){8-9}  

    &  {\bf FP} & {\bf FN}  & {\bf FP} & {\bf FN} & {\bf FP} & {\bf FN} & {\bf FP} & {\bf FN}  \\

  \midrule

  ATLASv2 & \UseMacro{ATLASv2_COMMANDER_Stage1_FP} & \UseMacro{ATLASv2_COMMANDER_Stage1_FN} & \UseMacro{ATLASv2_COMMANDER_Stage2_FP} & \UseMacro{ATLASv2_COMMANDER_Stage2_FN} & - & - & \UseMacro{ATLASv2_CBC_FP} & \UseMacro{ATLASv2_CBC_FN} \\

  \bottomrule
  \end{tabular}
\label{tab:evaluation:cbc_results}
\end{table}
}

\PP{\Sys vs. prior PIDS}
Most state-of-the-art heuristics-based PIDS \cite{holmes2019, nodoze2019, rapsheet2020} are evaluated on either proprietary datasets or datasets without persistence attacks. Hence it is not possible for us to make direct comparison with them. However, they would, by construction, fail at detecting persistence. Because a forward or backward tracing will not reach an event of interest in the next phase, if attackers break down the entire cyber-kill-chain into multiple phases like in Figure~\ref{fig:StealthinessByPersistence}. In fact, persistence attacks can be used to totally evade them. 

Therefore, we envisioned a comparison with three most recent state-of-the-art learning/anomaly-based PIDS 
KAIROS~\cite{KAIROS}, FLASH~\cite{FLASH} and MAGIC~\cite{MAGIC}, which all show superior performance over other learning-based PIDS like \cite{provdetector2020,unicorn2020,shadewatcher,PROGRAPHER,wang2022threatrace}.  However, none of these works outline detection results regarding to persistence attacks on each dataset. 
Each attack graph shown in the original papers was created from a subset of DARPA E3, DARPA E5, or DARPA OpTC dataset. None of those chosen subsets contains a true persistence attack. This is little surprising, as we already discussed above that public datasets often lack persistence traces. True persistence attacks only exist in DARPA E5 Fivedirections subset and DARPA OpTC day 2 subset. 

MAGIC is evaluated on DARPA E3 dataset, but not on more recent DARPA E5, OpTC or any datasets containing true persistence attacks. Both KAIROS and FLASH are evaluated on OpTC dataset. Although KAIROS is also evaluated on several DARPA E5 subsets, the Fivedirections subset is not included. KAIROS does not provide information needed to run on DARPA E5 Fivedirections. Both KAIROS and FLASH are a complicated system with many hyper-parameters and model configurations, it is challenging to ensure fair extension of their evaluation to E5 Fivedirections. Hence we took the pre-trained model weights for OpTC as provided by the authors of KAIROS and FLASH, and ran them only on the OpTC dataset and only on system events from the host containing a true persistence attack, respectively. 

Both KAIROS and FLASH perform attack detection at a finer granularity than previous PIDS like UNICORN. KAIROS takes system events as input, splits the entire time line into many time windows, and classifies each time window as benign or malicious, whereas FLASH can classify each node in the provenance graph as benign or malicious. KAIROS's detection result on OpTC shows that it has correctly classified the time window, in which persistence setup was conducted, as malicious, but wrongly classified the time window, in which the corresponding persistence execution happened, as benign. Similarly, FLASH produces a set of malicious nodes, which include the nodes related to persistence setup, but not the nodes responsible for persistence execution. As shown in Table~\ref{tab:evaluation:kairos_results}, \Sys, as a dedicated persistence detection system, requires 4× less memory resource than KAIROS and FLASH, while being more accurate and 315× faster than KAIROS, 156× faster than FLASH. Note that \Sys interfaces with Elasticsearch, a scalable and near real-time search engine, for rule matching, and performs provenance analytics only on system events related to persistence setup and execution.

\begin{table}
\centering
  \caption{Comparison of \Sys, KAIROS and FLASH on DARPA OpTC (Host 0501). \V = Detected, \xmark = Not Detected}
 \begin{threeparttable}
  \resizebox{0.48\textwidth}{!}{%
  \begin{tabular}{ccccc}
    \toprule
     & \makecell{Persistence\\Setup}  & \makecell{Persistence\\Execution}  & \makecell{Run Time\\(minute){*}} & \makecell{Mean Memory\\Consumption (GB)} \\
    \midrule
     \Sys & \V & \V   & 2 & 2.6\\
    \midrule
     KAIROS & \V & \xmark & 630 & 10.2\\
    \midrule
     FLASH & \V & \xmark & 312 & 9.1\\
    \bottomrule
  \end{tabular}}
    \begin{tablenotes}\footnotesize
   \item[*] From data processing to detection result.
   \end{tablenotes}
   \end{threeparttable}
  \label{tab:evaluation:kairos_results}
\end{table}

\subsection{Log Reduction via Expert-Guided Edges }
\label{ss:logReduction}
As discussed in Section~\ref{s:design}, with system logs generated by most popular logging frameworks we observe missing links due to the absence of IPC log events. However, by introducing expert-guided edges, we can cut the dependence on IPC events while still capable of detecting relevant attack steps.
During implementation of MITRE full emulation plans, we collect ALPC log events using the \quotes{NT Kernel Logger}~\cite{NTKernelLogger} ETW trace session. Table~\ref{tab:evaluation:logReduction} shows the log reduction rate of \Sys by employing expert-guided edges instead of relying on ALPC logs on each dataset.  
\begin{table}[!t]
\centering
  \caption{Log reduction rate of expert-guided edges}
  \resizebox{0.48\textwidth}{!}{%
  \begin{tabular}{cccc}
    \toprule
    \makecell{Dataset} & \makecell{Data Size}  & \makecell{ALPC Data Size} & \makecell{Reduction Rate} \\
    \midrule
     EP-APT29-1 & 32GB & 12GB &  38\%  \\
    \midrule
     EP-APT29-2 & 24GB & 5GB &  20\% \\
    \midrule
     EP-Sandworm-1 & 68GB & 36GB & 53\%  \\
    \bottomrule
  \end{tabular}}
  \label{tab:evaluation:logReduction}
\end{table}

\begin{figure*}[!t]
  \centering
  \includegraphics[width=0.9\linewidth]{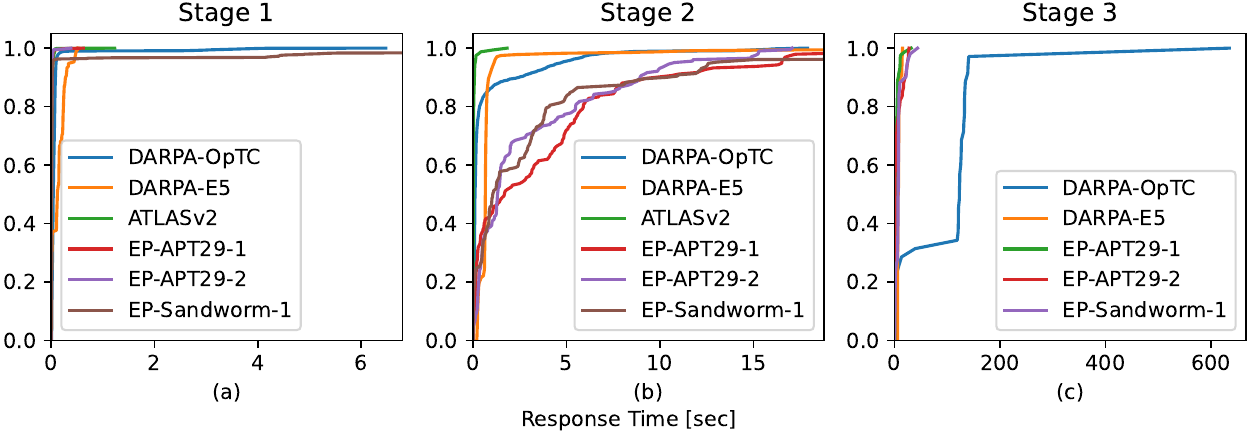}
  \caption{CDF of response time of \Sys}
  \label{fig:ResponseTime}
  \vspace{-2ex}
\end{figure*}

\begin{figure}[!t]
  \centering
  \includegraphics[width=0.95\linewidth]{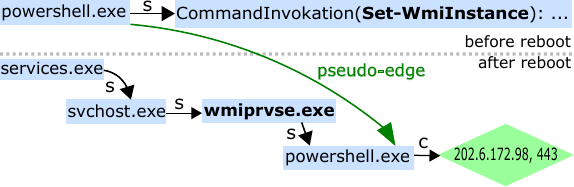}
  \caption{WMI persistence attack graph automatically generated by \Sys on DARPA OpTC dataset.}
  \label{fig:OptcPersistence}
\end{figure}

\subsection{Response Time \& Runtime Overhead}
\label{ss:ResponseTime}

We divide the response time of \Sys into three parts. Figure~\ref{fig:ResponseTime} presents the cumulative distribution function of response time of \Sys in its three stages, respectively. The stage 1 response time is measured per detection rule matching. Figure~\ref{fig:ResponseTime} (a) shows that it takes less than half a second to find alert events in an entire dataset for more than 95\% of detection rules. We measure the stage 2 response time  as the time to perform backward tracing on an indicative process with remote connection(s) while matching against persistence execution rules, persistence tables alignment and \Edge creation. As shown in Figure~\ref{fig:ResponseTime} (b), it takes less than 11 seconds to create a \Edge for over 90\% of all indicative processes. The stage 3 response time is the time for performing backward and forward tracing on each alert produced from stage 2 while checking for contextual indicators, and calculating the threat score. Figure~\ref{fig:ResponseTime} (c) shows that it takes less than 140 seconds to inspect each alert for over 95\% of all alerts in DARPA-OpTC dataset, and only less than 42 seconds for all alerts in other datasets. Memory overhead caused by running \Sys on each dataset is presented in Table~\ref{tab:evaluation:overhead}. The results are measured by using \texttt{mprof}~\cite{mprof}, which samples memory consumption every 100 ms.
\begin{table}
\centering
  \caption{Memory utilization (MB) of running \Sys}
  \begin{threeparttable}
  \resizebox{0.48\textwidth}{!}{%
  \begin{tabular}{ccccccc}
    \toprule
    & \makecell{DARPA-\\OpTC} & \makecell{DARPA-\\E5}  & \makecell{ATLASv2} & \makecell{EP-\\APT29-1} & \makecell{EP-\\APT29-2}  & \makecell{EP-\\Sandworm-1} \\
    \midrule
  max &  \UseMacro{DARPA_OpTC_Memory_Usage_max} & \UseMacro{DARPA_E5_Memory_Usage_max} & \UseMacro{ATLASv2_Memory_Usage_max} &  \UseMacro{APT29_Scenario1_Memory_Usage_max} &  \UseMacro{APT29_Scenario2_Memory_Usage_max} &  \UseMacro{Sandworm_Scenario1_Memory_Usage_max} \\
    \midrule
  mean  &  \UseMacro{DARPA_OpTC_Memory_Usage_mean} & \UseMacro{DARPA_E5_Memory_Usage_mean} & \UseMacro{ATLASv2_Memory_Usage_mean} &  \UseMacro{APT29_Scenario1_Memory_Usage_mean} &  \UseMacro{APT29_Scenario2_Memory_Usage_mean} &  \UseMacro{Sandworm_Scenario1_Memory_Usage_mean} \\
    \bottomrule
  \end{tabular}}
   \end{threeparttable}
  \label{tab:evaluation:overhead}
\end{table}

\subsection{Reconstructed Persistence Attack Graphs}
\label{ss:AttackGraph}
An example of reconstructed persistence attack graph from the public dataset OpTC is showed in the Figure~\ref{fig:OptcPersistence}. This figure depicts a true positive of T1546.003 (Event Triggered Execution: WMI Event Subscription). Like Figure~\ref{fig:APT29Persistence} and Figure~\ref{fig:FalsePositive} presented in Section~\ref{s:design}, the upper part of the figure shows the persistence setup graph, whereas the lower part shows the persistence execution graph. Both parts are connected via a \Edge.  This succinct attack graph exhibits that the attacker created a WMI instance during persistence setup. During persistence execution, the Powershell code for connecting back to the attacker is executed through the Windows system process \texttt{wmiprvse.exe}, when a specified event is triggered. We stress that the preciseness of this succinct attack graph containing the most critical information related to persistence can help even inexperienced security analysts achieve a speedy full attack investigation.

\section{Limitations \& Discussion}
\label{s:discussion}

\subsection{Evasion (Mimicry) Attacks}
As described in Section~\ref{s:ThreatModel}, \Sys is not designed to detect all persistence \mbox{(sub-)techniques}. Although \Sys is not tested to detect macOS-based persistence \mbox{(sub-)techniques}, we believe the principle is transferable. However, detecting cloud infrastructures-related persistence \mbox{(sub-)techniques} may require a different strategy with cloud infrastructures-specific conditions considered. Like all other provenance-based detection systems, which assume the OS integrity, \Sys is not able to detect pre-OS boot persistence attacks like Bootkit. However, none of these unaddressed persistence \mbox{(sub-)techniques} fall into the most misused top 10 persistence \mbox{(sub-)techniques}. That is, we believe this will have a limited impact on the practicality of \Sys. Further, \Sys is robust against evasion techniques proposed in~\cite{akul2023,Mukherjee2023}, as they specifically target anomaly-based PIDS that are based on path-based embedding or graph-based embedding. Heuristic-based systems like ours are more difficult to evade, as adding additional \quotes{camouflaging} events has little effect on these systems' detection results.

\subsection{Maintain and Extend \Sys}
One limitation of \Sys is that it relies on existing threat intelligence knowledge base (MITRE ATT\&CK), which is subject to expansion. That is, the detection rule base in stage 1 and the indicator list in stage 3 of \Sys need to be updated, if new attack techniques or behaviors emerge in the wild. Nonetheless, manually updating the rule base in stage 1 and the indicator list in stage 3 takes only a few minutes. Besides, we stress that unlike signature-based IDS relying on easily modifiable hard-coded strings, \Sys is based on characteristic attack behaviors and mostly uses immutable indicative strings in its stage 1.  In other words, \Sys depends on MITRE ATT\&CK Matrix and high-level behavior rules that are subject to change in a much slower pace. By tracking changes made in MITRE ATT\&CK Matrix, we find that MITRE updates the Matrix on a half-year basis. Our detection rule base and indicator list are based on a previous release (April 25, 2023). After inspecting the persistence tactic and techniques in the current release (April 23, 2024), we find that we do not need to update \Sys at all.

\subsection{Adaptability and Generality of \Sys}
\Sys is based on MITRE ATT\&CK Matrix, which is a general framework valued by organizations across the globe. As such, leading security vendors, on the one hand, continue contributing to this framework and, on the other hand, use this framework as a reference to develop detection rules/mechanisms.
Besides, we use popular standard instrumentation-free logging frameworks for collecting system logs. Our \Sys prototype interfaces with Elasticsearch, which is one of the most popular tools for event search and threat analysis among organizations. \Sys is tested on a typical Windows Domain network (including both Windows and Linux machines as being monitored clients) as required in the MITRE emulation plans. \Sys is deployed on a Linux machine functioning as a server that processes system logs shipped from client machines and performs attack detection. These characteristics, combined with its robust foundation in the MITRE ATT\&CK framework, establish \Sys as a versatile and comprehensive solution, readily deployable in a wide range of enterprise settings for effective persistence threat detection with minimal implementation effort.

\subsection{Completeness of \Sys}
\Sys's methodology, while illustrated through specific examples and MITRE techniques in the paper, embodies a comprehensive and generalizable framework for detecting persistence threats. The case-by-case analysis serves not merely as isolated instances but as representative samples of broader persistence threat patterns, showcasing \Sys's practicality across varied threat landscapes. This approach ensures that while the examples may appear specific in the paper, the underlying principles -- such as the segmentation into setup and execution phases -- are universally applicable. Such a strategy underlines \Sys's completeness, affirming its capability to address not just known scenarios but also to adapt and respond to emerging threats. \Sys's effectiveness is further validated through extensive evaluation on diverse datasets, showcasing its superior attack detection rates and graph completeness compared to state-of-the-art methods

\section{Related work}
\label{s:relwk}

In Section~\ref{s:intro}, we described the limitations of the existing threat detection system that \Sys addresses, and complement
the discussion on related work here:

\PP{Provenance-based IDS (PIDS)} Learning-based PIDS use machine learning to model benign behavior from provenance graphs. They alert on deviations during runtime. Early models, such as \cite{unicorn2020,PROGRAPHER}, detected anomalies at the graph level, complicating attack investigations. Newer models improve detection by focusing on time windows \cite{KAIROS}, edges \cite{shadewatcher}, or nodes \cite{wang2022threatrace,FLASH,MAGIC}. Despite advancements, these systems lack explanations for alerts, reducing interpretability. They also require extensive training data and slow down detection processes. Furthermore, they are vulnerable to concept drift and evasion attacks \cite{akul2023,Mukherjee2023}. Recent systems like \cite{KAIROS,FLASH,MAGIC} have addressed some issues but still fail to fully understand persistence attacks or include necessary contextual checks.

Heuristics-based PIDS, such as \cite{holmes2019,rapsheet2020}, apply detection signatures on provenance graphs to transform them into high-level APT stage graphs. These systems calculate threat scores based partly on the rarity of events and the correlation of APT stages within the APT graphs. However, they face difficulties in linking fragmented APT attack stages due to not recognizing the dual nature of persistence attacks, often resulting in incomplete attack graph reconstructions. Consequently, they rank disconnected graphs so low that true attacks frequently remain uninvestigated. Moreover, these systems require benign training data to assign rarity scores to events and filter false alarms. Additionally, RapSheet~\cite{rapsheet2020} is designed as an offline detection system, leading to significant delays in threat identification and investigation. Unlike these systems, \Sys excels in detecting persistence techniques by combining advanced detection rules with \Edges and expert-guided edges. This unique approach enables \Sys to effectively identify and investigate persistence threats in real-time, bridging the gaps that other PIDS fail to address.

\PP{Specialized Threat Detectors}
Specialized threat detectors focusing on single stages of APT attacks are well-documented in the literature. For instance, Ho et al.~\cite{ho2021hopper} introduced the Hopper system, and King and Huang~\cite{king2022euler} developed the Euler system, both targeting lateral movement detection using network logs. There are also specialized systems for detecting phishing emails~\cite{GrantHo2017}, data exfiltration~\cite{Ozery2024}, command and control (C2) activities~\cite{Carlos2020}, and ransomware~\cite{UNVEIL}. To the best of our knowledge, \Sys is the first specialized detector that focuses on persistence threats using provenance analytics, filling a critical gap in APT defense strategies.

\PP{Log Reduction Schemes} Numerous log reduction systems have been proposed recently, such as~\cite{inam2022sok,Xu2016,elise,tang2018nodemerge,hossain2018}. Unlike these systems, \Sys specifically aims to reduce the reliance on IPC logs to enhance persistence detection, a focus not typically addressed by existing log reduction schemes. These systems, while complementary to \Sys, can also be integrated to further decrease the size of audit logs.

\section{Conclusion}
\label{s:conclusion}
In this paper, we introduce \Sys, a novel system dedicated to detecting persistence attacks. Distinctively, \Sys leverages provenance analytics, moving beyond the basic detection rules traditionally used. This approach not only significantly reduces false alarms but also notably enhances accuracy in identifying genuine persistence attacks. Our evaluations, conducted on both public datasets and datasets derived from rigorously executed MITRE emulation plans, demonstrate \Sys's superiority over state-of-the-art detection methods. Furthermore, \Sys incurs low runtime overhead, making it a valuable addition to the suite of threat detectors in enterprise settings.

\balance

\section*{References}
\printbibliography[heading=none]

\end{document}

\typeout{get arXiv to do 4 passes: Label(s) may have changed. Rerun}